\documentclass[english,12pt]{article}
\usepackage{graphicx}
\usepackage{subfig}
\usepackage{amsmath}
\usepackage{amssymb}
\usepackage{bm}
\usepackage{slashed}
\usepackage{dcolumn}
\usepackage{epsfig}
\usepackage{times}
\usepackage{cite}
\usepackage{afterpage}
\usepackage{color}
\usepackage[T1]{fontenc}
\usepackage[latin9]{inputenc}
\usepackage{babel}
\usepackage{lipsum}
\usepackage{ulem} 
\usepackage{url}

\allowdisplaybreaks[1]
\newcommand{\lsim}{\raisebox{-4pt}{$\,\stackrel{\textstyle
                                                         <}{\sim}\,$}}

\newcommand{\nn}{\nonumber}
\newcommand{\be}{\begin{equation}}
\newcommand{\ee}{\end{equation}}
\newcommand{\ba}{\begin{eqnarray}}
\newcommand{\ea}{\end{eqnarray}}
\newcommand{\req}[1]{(\ref{#1})}
\def\={\,=\,}
\newcommand{\ci}[1]{\cite{#1}}

\newcommand{\bea}{\begin{eqnarray}}
\newcommand{\eea}{\end{eqnarray}}
\newcommand{\beq}{\begin{equation}}
\newcommand{\eeq}{\end{equation}}

\newcommand{\gep}{$G_E^p$}

\newcommand{\KLL}{$K_{_{LL}}$}

\newcommand{\PR}{{ Phys. Rev. }}
\newcommand{\PRL}{{ Phys. Rev. Lett. }}
\newcommand{\PL}{{ Phys. Lett. }}
\newcommand{\NP}{{ Nucl. Phys. }}

\newcommand{\etal}{{\em et al.}}

\def\lambdabar{\lambda\kern-1ex\raise0.65ex\hbox{-}}

\def\gev{~{\rm GeV}}

\def\ale{\alpha_{\rm em}}

\def\vb0{{\bf b}_0}

\newcommand{\wf}{wave function}

\def\sh{\hat{s}}
\def\uh{\hat{u}}
\def\th{\hat{t}}
\usepackage[margin=1in]{geometry}
\begin{document}
\hfill{\bf Date:} {\today}
\begin{center}
{\huge\bf Workshop on High-Intensity Photon Sources \\
	(HIPS2017) \\
	Mini-Proceedings} 
\end{center}
\hspace{0.1in}
\begin{center}
{\large 6th - 7th February, 2017}
{\large Catholic University of America, Washington 
	, DC, U.S.A.}
\end{center}
\begin{center}
S.~Ali,
L.~Allison,
M.~Amaryan,
R.~Beminiwattha,
A.~Camsonne,
M.~Carmignotto,
D.~Day,
P.~Degtiarenko,
D.~Dutta,
R.~Ent,
J.~L.~Goity,
D.~Hamilton,
O.~Hen,
T.~Horn,
C.~Hyde,
G.~Kalicy,
D.~Keller,
C.~Keppel,
C.~Kim,
E.~Kinney,
P.~Kroll,
A.~Larionov,
S.~Liuti,
M.~Mai,
A.~Mkrtchyan,
H.~Mkrtchyan,
C.~Munoz-Camacho,
J.~Napolitano,
G.~Niculescu,
M.~Patsyuk,
G.~Perera,
H.~Rashad,
J.~Roche,
M.~Sargsian,
S.~Sirca,
I.~Strakovsky,
M.~Strikman,
V.~Tadevosyan,
R.~Trotta,
R.~Uniyal,
A.H.~Vargas,
B.~Wojtsekhowski,
, and
J.~Zhang
\end{center}
\begin{center}
\textbf{Editors}: T.~Horn, C.~Keppel, C.~Munoz-Camacho, and I.~Strakovsky
\end{center}

\noindent

\begin{center}{\large\bf Abstract}\end{center}

This workshop aimed at producing an optimized photon source concept with potential increase of scientific output at Jefferson Lab, and at refining the science for hadron physics experiments benefitting from such a high-intensity photon source. The workshop brought together the communities directly using such sources for photo-production experiments, or for conversion into $K_L$ beams. The combination of high precision calorimetry and high intensity photon sources greatly enhances scientific benefit to (deep) exclusive processes like wide-angle and time-like Compton scattering. Potential prospects of such a high-intensity source with modern polarized targets were also discussed. The availability of $K_L$ beams would open new avenues for hadron spectroscopy, for example for the investigation of "missing" hyperon resonances, with potential impact on QCD thermodynamics and on freeze-out both in heavy ion collisions and in the early universe.


\newpage
\tableofcontents

\newpage
\pagenumbering{arabic}
\setcounter{page}{1}
\section{Preface}
\halign{#\hfil&\quad#\hfil\cr
}

\begin{enumerate}
\item \textbf{Preface}

This volume presents the mini-proceedings of the workshop on "High-Intensity Photon Sources" which was held at The Catholic University of America in Washington DC, USA, February 6-7, 2017. 

There has been recent discussion of a photon source with large gain in figure-of-merit to be used with both dynamically polarized targets to measure processes such as wide-angle and timelike Compton Scattering (WACS and TCS), and potentially to convert into a $K_L$ beam for spectroscopy experiments. PAC43 and PAC44 at Jefferson Lab have seen a few proposals and several LOIs related to these photoproduction topics. The possible photon source would give a gain in figure-or-merit of a factor of 30 for some experiments. For processes such as wide-angle compton scattering and timelike compton scattering the high intensity photon source could be coupled with high-precision calorimetry. For the latter, plans are discussed in all four Halls. 

This workshop brought together the theory and experiment communities using high precision calorimetry and high intensity photon sources, with the aim
to consider the high-level science goals to be achieved by WACS and TCS processes, The workshop also included discussions of converting a
high intensity photon source into $K_L$ beam. The emphasis was on new avenues in hadron structure studies that could contribute to the existing
program at JLab. By exploring new avenues for neutral kaon beams, it may also contribute to the hadron spectroscopy program.

The workshop consisted of 19 talks and discussion and summary sessions. We had about 45 participants from 20 institutions, some of them international. More than 30\% of the participants were young scientists or students showing that this field is interesting and has growing importance.

These mini-proceedings summarize 17 of the 19 talks presented at the
workshop. The actual talks in pdf format may be found on the web page 
of the workshop: \\
https://www.jlab.org/conferences/HIPS2017/ .

Finally we would like to take this opportunity to thank all speakers,
session chairs, secretaries, and all participants for making this 
workshop a real success.

\newpage
\item \textbf{Acknowledgments}

The workshop could not have been done without dedicated work of many people. We would like to thank the Catholic University of America and the Vitreous State Laboratory, in particular Briana Zakszeski, Tom Santoliquido and Ian Pegg, for local support hosting this event, the Staff Services and Electronic Media groups at JLab for help with the workshop web page, JSA Initiatives funds, JLab management, especially Robert~McKeown for their help and encouragement to organize this workshop. 

\vskip 1cm
\leftline {Washington DC, February 2017.}

\rightline {T.~Horn}
\rightline {C.~Keppel}
\rightline {C.~Munoz-Camacho}
\rightline {I.~Strakovsky}

\end{enumerate}

%

\newpage
\section{Summaries of Talks}


\subsection{A Pure Photon Beam for  Use with Solid Polarized Targets}
\addtocontents{toc}{\hspace{2cm}{\sl D.~Day}\par}
\setcounter{figure}{0}
\setcounter{table}{0}
\setcounter{equation}{0}
\setcounter{footnote}{0}
\halign{#\hfil&\quad#\hfil\cr
\large{Donal Day\footnote{Email:  dbd@virginia.edu}}\cr
\large{Dustin Keller \footnote{Email:  dustin@jlab.org}}\cr
\large{Darshana Perea \footnote{Email: darshana@jlab.org}}\cr
\large{Jixie Zhang \footnote{Email:  jixie@jlab.org}}\cr
\textit{University of Virginia}\cr
\textit{Physics Department}\cr
\textit{Charlottesville, Virginia, U.S.A.}\cr}

\begin{abstract}
The modest luminosity of solid polarized targets with charged particles limits their use in real photon experiments  where the mixed photon-electron beam produced at a radiator passes through the target. A pure photon beam will increase the overall figure of merit when used with a solid polarized target by a factor of 30 or more. This increase is in part due to a higher average polarization of the target associated with minimizing the radiation damage to the target and the depolarization the damage causes, along with an increase in the useful photon flux. A separated function source, where the electron radiator, dipole and collimator are divorced from the beam dump is described, exposing its attractive aspects. Work necessary to perfect this approach is ongoing. 
\end{abstract}

\begin{enumerate}
\item \textbf{Introduction}
 Significant progress has been made over the last decade in our understanding of exclusive reactions in the hard scattering regime. Jefferson Lab data on elastic electron scattering and Compton scattering has made much of this progress possible. Specifically,  these are the recoil polarization measurements of \gep\, E93-027, E04-108 and E99-007, and the Real Compton Scattering (RCS) experiments E99-114 and E-07-002. The \gep~measurements~\cite{Jones:1999rz,Gayou:2001qd,Puckett:2010ac} found that the ratio of $F_2$ and $F_1$, scaled by $Q^2$ demands a redrafting of one of the canons of pQCD, namely hadron helicity conservation. Results from the RCS measurements~\cite{Hamilton:2004fq,Fanelli:2015eoa} are that the longitudinal polarization transfer \KLL~is large and positive, contrary to the pQCD predictions for \KLL.  Together, these experiments obligate us to suggest that pQCD can not be applied to exclusive processes at energy scales of 5-10 GeV.

Polarized RCS from polarized targets, for example, Jefferson Lab's E12-14-006\cite{Keller2014}, can contribute to the study of hard exclusive reactions, complimenting the recoil polarization measurements from unpolarized targets. Until recently it was thought that the kinematic reach and precision of polarized target experiments would lag due to their inherent luminosity limitations when used with the mixed electron-photon beam produced by the bremsstrahlung radiator. This limitation can be largely lifted by sweeping the bremsstrahlung electrons away  (to a local low-power beam dump) so that only a pure photon beam reaches the polarized target. Recent proposals ~\cite{BW2015}\cite{Keller2016} are contingent upon a pure photon beam, relegating a mixed photon-electron beam as a less than optimal approach.

\item \textbf{Pure Photon Source}
Figure~\ref{fig:layout} provides  the setup which has been simulated in G4beamline and Geant4. The electron beam comes in from the left striking a copper radiator. A 2m-long 2T dipole, tilted at 5$^\circ$, follows the radiator and deflects the electron beam by approximately 8$^\circ$ (for 8.8 GeV) such that it passes through a gap between the bottom of the polarized target vacuum can and the pivot post (in Hall C). The electron beam drifts to a 25kW dump on the floor. The photon beam passes through a combined collimator/absorber to the target and then to the hall dump. The benefit of this approach can be immediately recognized in that the beam dump is moved far downstream facilitating massive shielding while avoiding a dump adjacent to the target where space is limited and where regular access is required.
 
 The above, conceptually simple, it is complex in practice due to the energy and angular spread of the electron beam out the radiator, the need to collimate the photon beam, potential backgrounds generated along the beam line to the local dump, and dump and collimator shielding requirements.
 
 Energy loss to photons in the radiator creates a significant energy spread in the deflected electron beam and multiple scattering widens it transverse distribution as can be seen in Figure~\ref{fig:beamspread} where we show the electron beam distribution just after the dipole. There it can be seen that the vertical extent of the beam is quite large. It is not possible to drift such a beam downstream to the  local dump. A collimator/absorber to absorb these off-energy electrons will be installed just downstream of the dipole and serves two purposes. One is to absorb the off-energy electrons and the other is to collimate the photon beam.  The collimator design is based on the PREX/CREX one for Hall A and would absorb 2.5 - 3 kW. This is an area of still active study. It may well be beneficial from a radiological background standpoint to increase further the absorption in the collimator. The Moller experiment plans for  4~kW collimator.
 
 \begin{figure}[htbp] 
    \centering
    \includegraphics[width=0.5\textwidth]{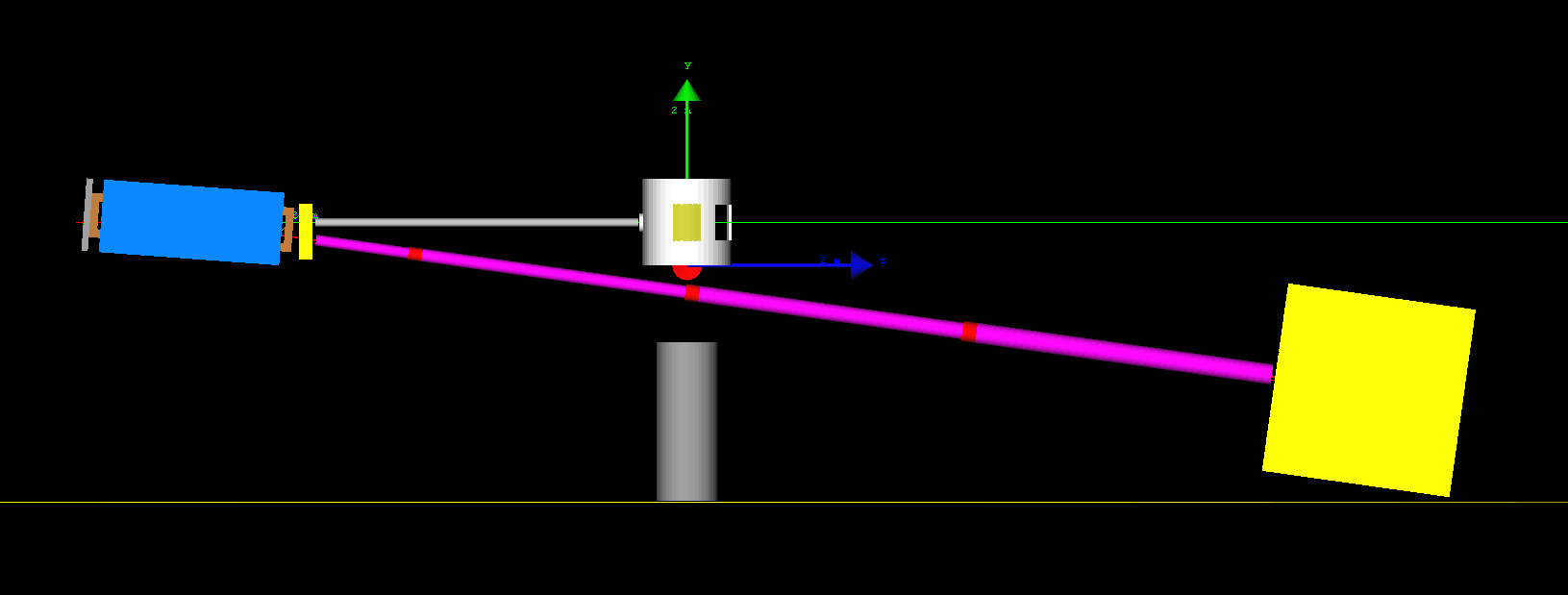}  
    \caption{Photon beam setup. See text for description. \label{fig:layout}}
 \end{figure}
Table~\ref{tab:kWatt} presents the energy deposition in the different elements of the beam line for both 10\% and 6\% radiators. 
  \begin{figure}[!htbp] 
    \centering
    \includegraphics[width=\columnwidth]{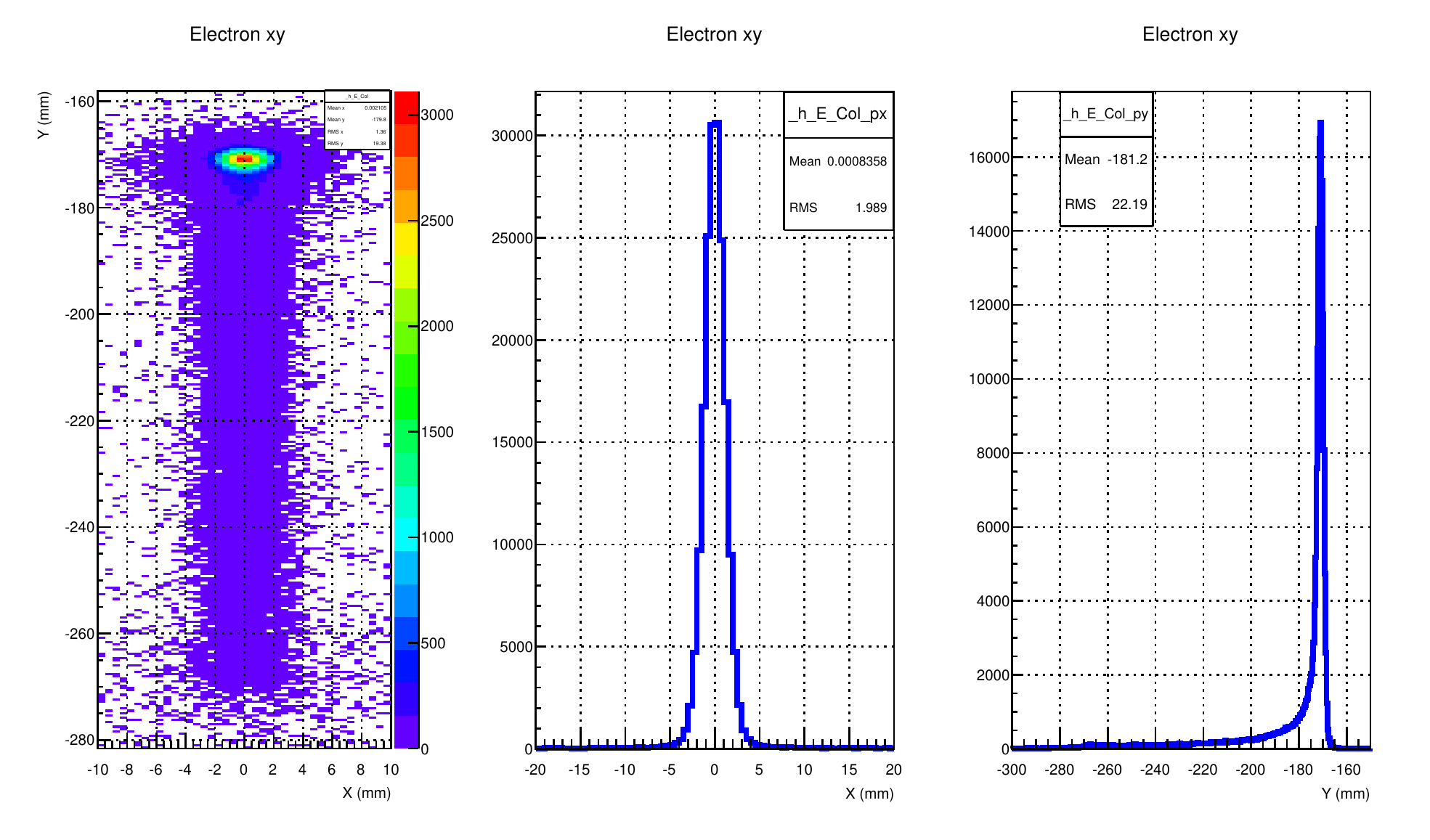}
    \caption{Distribution of electrons exiting dipole. Energy loss and multiple scattering spreads the electron beam and widens its transverse distribution. A collimator/absorber at the exit of the dipole  will reduce the vertical extent to a few cm.\label{fig:beamspread}}
 \end{figure}

 Our studies to date have not fully exploited the possible parameter space available here: maximum field in a resistive dipole, distance from the dipole (radiator) to the target can, dipole length and dump location. These studies are in progress  as are continued studies of the radiation profile using FLUKA, both prompt and accumulated, at various locations in the hall with multiple shielding possibilities\cite{JixieZ2017}. In addition, collaborative efforts will  compare this concept to the Compact Photon Source\cite{GNic2017} so as to develop a feasible, flexible and robust source of pure polarized photons for future polarized target experiments.

\begin{table}[!ht]
\caption{Energy Deposition}
\centering
\begin{tabular}{c c  c}
\hline\hline
Component & (kW/$\mu$A) 10\% r.l. &(kW/$\mu$A) 6\% r.l.\\
\hline
Radiator&  0.00191 & 0.00112 \\
Dipole magnet &
0.184&
0.106\\
Collimator&
0.784&
0.456 \\
Photon beam pipe &
0.00143 &
0.00097\\ 
Electron beam pipe &
0.018&
0.012\\
Hall Dump&
0.418&
0.329\\
Local Dump&
6.65&
7.46\\
Flux at the Target&
$1.42(10)^{11}$&
$1.16(10)^{11}$\\
\hline
\end{tabular}\caption{Results of a simulation of an 8.8 GeV  beam on 6\% and 10\% radiators. The photon flux is that within a 2mm spot at the target cell and for photon energies $ 0.5 < E_\gamma/E_{beam} < 0.95$.}
\label{tab:kWatt}
\end{table}

\item \textbf{Acknowledgments} This work was supported by DOE Grant Number DE-FG02-96ER40950.

\end{enumerate}


\newpage
\subsection{The concept of the Hermetic Compact Photon Source}
\addtocontents{toc}{\hspace{2cm}{\sl G.~Niculescu}\par}
\setcounter{figure}{0}
\setcounter{table}{0}
\setcounter{equation}{0}
\setcounter{footnote}{0}
\halign{#\hfil&\quad#\hfil\cr
\large{Gabriel Niculescu \footnote{Email:  gabriel@jlab.org}}\cr
\textit{James Madison University}\cr
\textit{Physics Department}\cr
\textit{U.S.A.}\cr}

\begin{abstract}
The concept and projected performance of a Hermetic Compact Photon are presented. As designed the HCPS represents the optimal solution, minimizing cost, volume, and weight while maximizing the flux for all experiments seeking to explore polarization observables using photon beams at Jefferson Lab. 
\end{abstract}

\begin{enumerate}
\item \textbf{Introduction}
Spurred by the promise of accessing Generalized Parton Distributions the interest in exploring photon--induced reactions has within the Jefferson Lab scientific community over the last few years. While mixed electron--photon beams are still the tool of choice for measuring unpolarized cross--sections, (double) polarization observables such as those obtained in a Wide Angle Compton Scattering experiment\ \cite{bogdan}, due to the nature of the target and the analysis details, call for a pure photon beam. The Hermetic Compact Photon Source (HCPS) described below provides the optimal solution for this problem.

\item \textbf{The Hermetic Compact Photon Source}

Bremsstrahlung (i.e. the use of a radiator) is the preferred way of producing photon beam starting with the primary JLab electron beam. This produces a mixed electron/photon beam. To get a pure photon beam then the remaining electrons have to be bent away from the target and disposed off, usually by the means of a dump. If the thickness of the original radiator is small (say $\sim 10^{-4}$ radiation lengths) then it will produce a minimal disturbance to the electron beam while and one can consider bending this beam around the target and directing it to a dump. However, only a very low intensity beam will be produced, an even smaller portion of which would be high enough in energy to be of interest for the kind of experiments one envisions (WACS, single pion photoproduction, etc.). 

Increasing the thickness of the radiator to about 10\% will provide the increase in intensity one needs. However, this benefit comes at the expense of a much more disturbed electron beam, both in terms of the electron energy distribution and divergence. In practical terms this means that the further from the radiator one gets, the larger the envelope of the residual electron beam becomes. Dipole magnetic fields used for bending the beam (for example to avoid the target) will only increase the size of the beam. Correspondingly, the beam pipe needed to contain this degraded beam and the aperture of the dump will have to be very large (the further from the original radiator, the larger), increasing the complexity, volume, and price tag of this approach. Assuming that they can be positioned close enough to the radiator, quadrupoles can be used to re--focus the residual beam which will help alleviate the beam pipe size requirements, at the cost of building and supporting said magnets. The $\sim$10\%
 radiator assumed in this scenario will be a large source of radiation in itself. A relatively straightforward calculation\ \cite{bogdan}, backed by our GEANT4 simulation, estimates that about 25~\% of the primary electron beam power is lost in the radiator, thus the region will require substantial shielding.
 
Given the technical and increased cost problems associated with the concept of transporting the residual electron beam to a dump a more sensible approach is to make the dump as close as reasonably achievable to the radiator. Ideally one would dump all of the residual beam at the radiator point and then build a spherically enclosure engineered to shield the environment from the resulting radiation. As one still needs to keep the radiator relatively thin so as to minimize the multiple scattering inside the radiator, and the logistics of building and supporting a multi--ton sphere are challenging, the Hermetic Compact Photon Source we developed achieves the next best thing: a device that accepts the primary electron beam through a small aperture, converts $\sim$10~\% of it into a photon beam, and acts as a beam dump for the residual electron beam. The whole device is enclosed in a cubical volume as shown in Figure\ \ref{fig:hcps},

\begin{figure}[!h]
\begin{center}
\includegraphics[width=12cm]{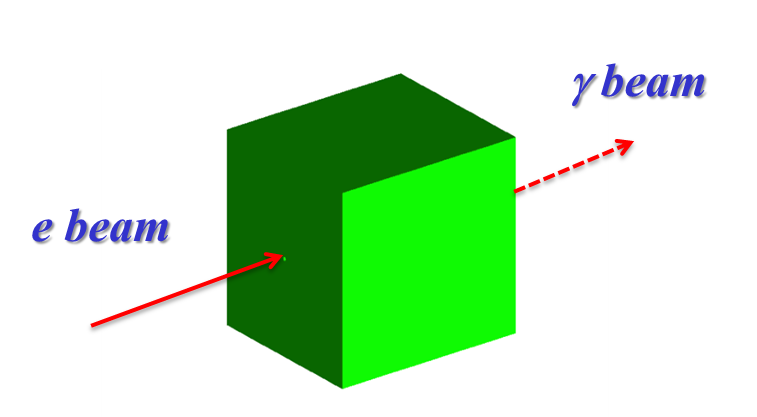}
\end{center}
\caption{Illustration of the Hermetic Compact Photon Source.}
\label{fig:hcps}
\end{figure}

To achieve this goal the HCPS uses a roughly cubical ($\sim$0.125~$m^3$) normal conducting magnet (see Figure\ \ref{fig:hcps2}) with a small bore and a field of $\sim$~2.5~T. This field will bend the residual beam into the magnet itself. Thus the magnet will act a beam dump. Layers of shielding around the magnet ensure that the resulting radiation is contained to acceptable levels. Analytic calculations as well as GEANT4 simulations of this setup show that the photon beam produced is very small, 0.9~mm diameter (Figure\ \ref{fig:hcps3}) on a target located $\sim$2~m away from the radiator - which is crucial for data analysis of WACS experiments where one needs to disentangle the Compton scattering events from $\pi^0$ events. Assuming a $1.2\mu A$ electron beam current at 8.8~GeV the HCPS should be able to provide $\sim 10^{11} - 10^{12}$ photons.

\begin{figure}[h]
\begin{center}
\includegraphics[width=12cm]{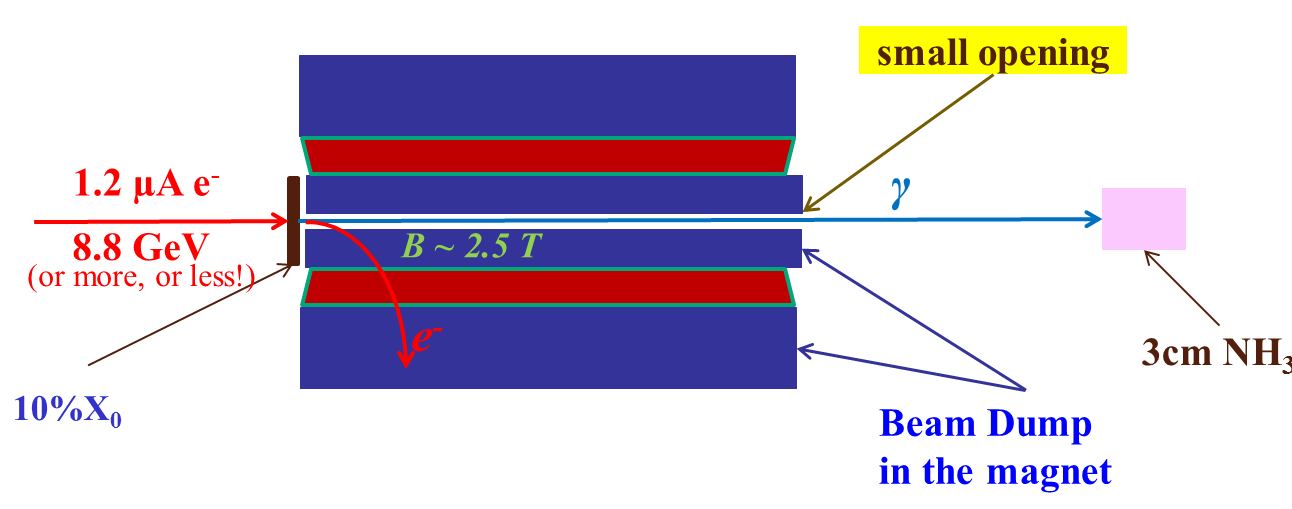}
\end{center}
\caption{Schematic of the HCPS magnet. Shielding surrounding the magnet on all sides are not shown.}
\label{fig:hcps2}
\end{figure}

Extensive simulation\ \cite{hcps_report}  of this photon source was carried out in GEANT4. This aimed and achieved to:

\begin{itemize}
\item{Determine the size and layering of the shielding around the magnet. A succession of concrete and iron blocks seems to achieve this goal. All the inside of the magnet is filled with non--magnetic inserts (Cu, CuW alloy). Further refinements of the shielding scheme including the potential use of boron--enriched concrete, W sheets, Pb sheets, high--density poly, etc. are ongoing.}
\item{Determine the radiation level on the polarized target electronics.}
\item{Determine the radiation level on the magnet coils and based on these results identify radiation hardened materials\protect \cite{radhard, jparc_magnets} that might be used in building the coils.}
\item{Determine the radiation level immediately next to the HCPS as well as at the experimental hall boundary.}
\end{itemize}

\begin{figure}[h]
\begin{center}
\includegraphics[width=12cm]{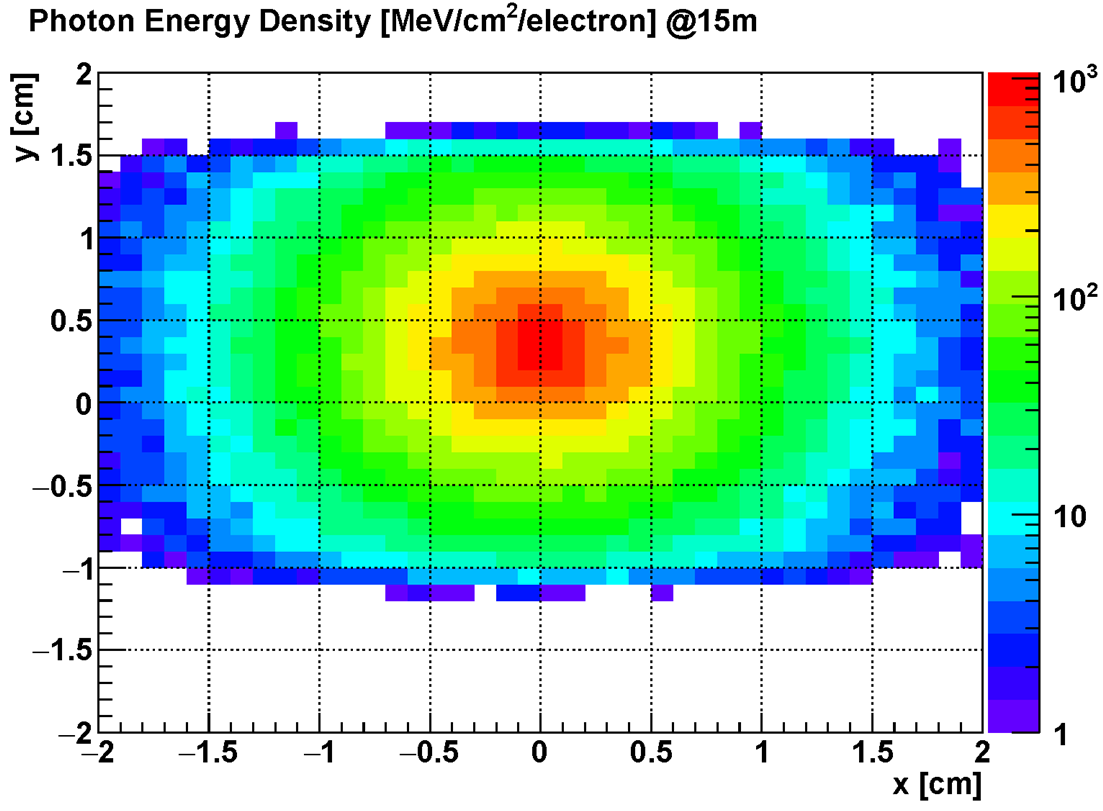}
\end{center}
\caption{HCPS photon beam profile. The beam spot size is $\sim$0.9~mm in diameter.}
\label{fig:hcps3}
\end{figure}

Currently we are in the process of refining the size and composition of the shielding in an attempt to reduce the mass of the HCPS and (potentially) reduce cost. In order to achieve this purpose extensive GEANT4 simulations are underway. In parallel we are studying the activation of the HCPS materials using FLUKA. Early results\ \cite{michael} indicate that the dose equivalent next to the HCPS, after a 1000~hour--long experiment using 1.2~$\mu A$, 8.8~GeV electron beam is at or below 1~mrem/h, as computed one hour after the primary beam is off.

\item \textbf{Conclusion}

The HCPS device described above manages to convert a portion of the primary JLab electron beam into a highly focused (untagged) photon beam. The source is {\bf H}ermetic as it contains all the initial energy not converted into the photon beam and keeps the radiation and activation to acceptable levels. It is {\bf C}ompact as it arguably provides the smallest possible volume that can be built with current materials and technology. Lastly, this {\bf P}hoton {\bf S}ource is ideally suitable for carrying out double polarization experiments such as, but not limited to, WACS. As designed, the HCPS is capable of handling the highest JLab beam energies. It is also scalable to lower energies, which might be of interest for medical and industry applications.

%

\end{enumerate}


\newpage
\subsection{The handbag approach to wide-angle Compton Scattering}
\addtocontents{toc}{\hspace{2cm}{\sl P.~Kroll}\par}
\setcounter{figure}{0}
\setcounter{table}{0}
\setcounter{equation}{0}
\setcounter{footnote}{0}
\halign{#\hfil&\quad#\hfil\cr
\large{Peter Kroll\footnote{Email:  pkroll@uni-wuppertal.de}}\cr
\textit{Universit\"at Wuppertal}\cr
\textit{Fachbereich Physik}\cr
\textit{D-42097, Wuppertal, Germany}\cr}

\begin{abstract}
In this talk I reviewed briefly the present status of the handbag approach
to wide-angle Compton scattering (WACS).
\end{abstract}

\begin{enumerate}
\item \textbf{Introduction}
In \cite{DFJK1} the WACS amplitudes have been derived for
Mandelstam variables $s, -t$ and $-u$ that are large as compared to $\Lambda^2$ where
$\Lambda$ is a typical hadronic scale of order $1\,\gev$. Under the assumption that the
soft hadronic \wf{} occurring in the Fock decomposition of the proton state, is dominated
by parton virtualities $|k_i^2|\lsim \Lambda^2$ and by intrinsic transverse momenta that
satisfy $k_{\perp i}^2/x_i\lsim \Lambda^2$, the light-cone helicity amplitudes for WACS,
${\cal M}$, factorize in amplitudes, ${\cal H}$, for Compton scattering off massless
quarks and in form factors which represent $1/x$-moments of zero-skewness generalized
parton distributions (GPDs)
\begin{equation}
R_V^a(t)= \int_{-1}^1 \frac{dx}{x} H^a(x,0,t)
\end{equation}
where $a$ denotes the quark flavor. The full form factor is
\begin{equation}
R_V = \sum_a e_a^2 R_V^a(t)
\end{equation}
The GPD ${\widetilde H}(E)$ is related to the axial (tensor) form factor $R_A (R_T)$ in an 
analogous fashion.  The WACS amplitudes read \cite{DFJK1,HKM}
\begin{align}
{\cal M}_{\mu'+,\mu +}(s,t)&=& 2\pi\ale \Big[{\cal H}_{\mu'+,\mu +}(\sh,\th)
       \big(R_V(t)+R_A(t)\big) \nn\\
            &&  + {\cal H}_{\mu'-,\mu -}(\sh,\th) \big(R_V(t)-R_A(t)\big)\nn\\
{\cal M}_{\mu'-,\mu +}(s,t)&=& \pi\ale\frac{\sqrt{-t}}{m} \Big[{\cal H}_{\mu'+,\mu +}(\sh,\th)
                 + {\cal H}_{\mu'+,\mu +}(\sh,\th)\Big]\,R_T(t)\,.
\label{eq:amplitudes}
\end{align}
The subprocess amplitudes have been calculated to order $\alpha_s$ \cite{HKM}. The large
$-t$ GPDs at zero skewness have been extracted in an analysis of the nucleon form factors
exploiting the sum rules \cite{DFJK4,DK13}. The Compton form factors can be evaluated
from these GPDs and, hence, the WACS cross section can be predicted free of parameters. The results
agree reasonably well with experiment. 

Predictions for spin-dependent observables have also been given \cite{HKM}. Some care 
is however necessary, in particular for a kinematical region where the Mandelstam variables
are not much larger than $\Lambda^2$. Various corrections have to be considered and a
quality assessment of the Compton form factors is necessary:
\begin{itemize}
\item The amplitudes \req{eq:amplitudes} are given for light-cone helicities. For a 
      comparison with experiment standard helicities are more convenient. In \ci{diehl01}
      the transformation from one basis to the other one is given. The admixtures of amplitudes
      with opposite proton helicity is under control of the parameter 
      $\eta=2m\sqrt{-t}/(s+\sqrt{-us})$. For the massless photons both light-cone and 
      standard helicities fall together.
\item The matching of the variables $s, t, u$ with the corresponding ones for the subprocess, 
      $\sh, \th, \uh$, is another source of uncertainty \cite{DFHK03}. For $s$ less than 
      $10\,\gev^2$ these uncertainties are large. 
\item The vector form factor $R_V$ is rather well determined due to the precise
      data on the magnetic form factor of the proton measured in a large range of $t$ which 
      dominate $R_V$ at large $-t$. The tensor form factor is less accurately  known
      since it is sensitive to all four electromagnetic form factors of the nucleon \cite{DK13}.
      The neutron form factors are only measured up to $4\,\gev^2$ as yet. The available
      experimental information on the axial form factor, $F_A$, at large $-t$ from which 
      in principle $\widetilde H$ is determined, is very limited. There are only dipole 
      parametrizations of $F_A$. This poor information prevents an analysis of $\widetilde H$ 
      like that one performed for $H$ and $E$. Several examples of parametrizations for     
      $\widetilde H$ have been presented which all agree with the data on $F_A$ within errors
      but lead to substantially different results for $R_A$ and, hence, for different 
      predictions for the spin correlation parameter $K_{LL}=A_{LL}$. In turn  from a 
      measurement of $K_{LL}$ or $A_{LL}$  at sufficiently large $-t$ and $-u$ one may 
      extract $R_A(t)$ and use these results as an additional constraint in the analysis of 
      $\widetilde H$ at large $-t$. Even if we would have new and better data on $F_A$, 
      e.g.\ from the FNAL MINERVA experiment, additional data on $R_A$ will help in the flavor
      decomposition of $\widetilde H$. Knowledge of the large $-t$ behavior of $\widetilde H$
      is of interest for investigating the transverse distribution of longitudinally polarized
      quarks in the proton.
\end{itemize}


\end{enumerate}


\newpage
\subsection{Photon Beam Requirements for Wide-Angle Compton Scattering}
\addtocontents{toc}{\hspace{2cm}{\sl D.J.~Hamilton}\par}
\setcounter{figure}{0}
\setcounter{table}{0}
\setcounter{equation}{0}
\setcounter{footnote}{0}
\halign{#\hfil&\quad#\hfil\cr
\large{David J. Hamilton\footnote{Email:  david.j.hamilton@glasgow.ac.uk}}\cr
\textit{University of Glasgow}\cr
\textit{SUPA School of Physics and Astronomy}\cr
\textit{Glasgow G12 8QQ, UK}\cr}

\begin{abstract}
  The ongoing efforts to develop of a high-intensity photon source at
  Jefferson Lab are partly driven by future measurements of the
  initial-state helicity observable $A_{LL}$ in wide-angle Compton
  scattering. While there are many technical challenges related to the
  development of such a source, the focus in the present report is on
  the impact of the photon beam parameters on future WACS
  measurements.
\end{abstract}

\begin{enumerate}
\item \textbf{Introduction}
The 6 GeV era at Jefferson Lab included two successful experiments on
Wide-Angle Compton Scattering (WACS) in Halls A~\cite{ham05, dan07}
and C~\cite{fan15}.  These experiments involved measurements of both
polarized and unpolarized observables for WACS over a broad region of
the kinematic range that was accessible. The dramatic improvement in
statistical precision compared with previous measurements~\cite{shu79}
was achieved primarily through the utilization of a high-intensity
mixed electron-photon beam and detector systems with high-rate
capability.

The experiments employed a proton spectrometer with good momentum and
angular resolution to detect the recoil proton (HRS or HMS) and a
highly segmented electromagnetic calorimeter to detect the scattered
photon. The analysis technique then relies on utilization of the
two-body kinematic correlation between the final-state particles,
which is dominated by events from three main reaction channels:
\begin{itemize}
\item $\gamma + p \rightarrow \gamma + p$;
\item $\gamma + p \rightarrow \gamma + \pi^0$;
\item $e + p \rightarrow e + p~(+~ep\gamma)$.
\end{itemize}
One then extracts observables from the difference distribution of
measured versus predicted (from the proton spectrometer) calorimeter
hit position ($\delta_x$ and $\delta_y$). A plot of a typical
distribution is shown in Fig.~\ref{fig:dy}, where the WACS events are
kinematically indistinguishable from background events corresponding
to both neutral pion and $ep\gamma$ reactions.

\begin{figure}
  \begin{center}
    \includegraphics[width=0.45\textwidth]{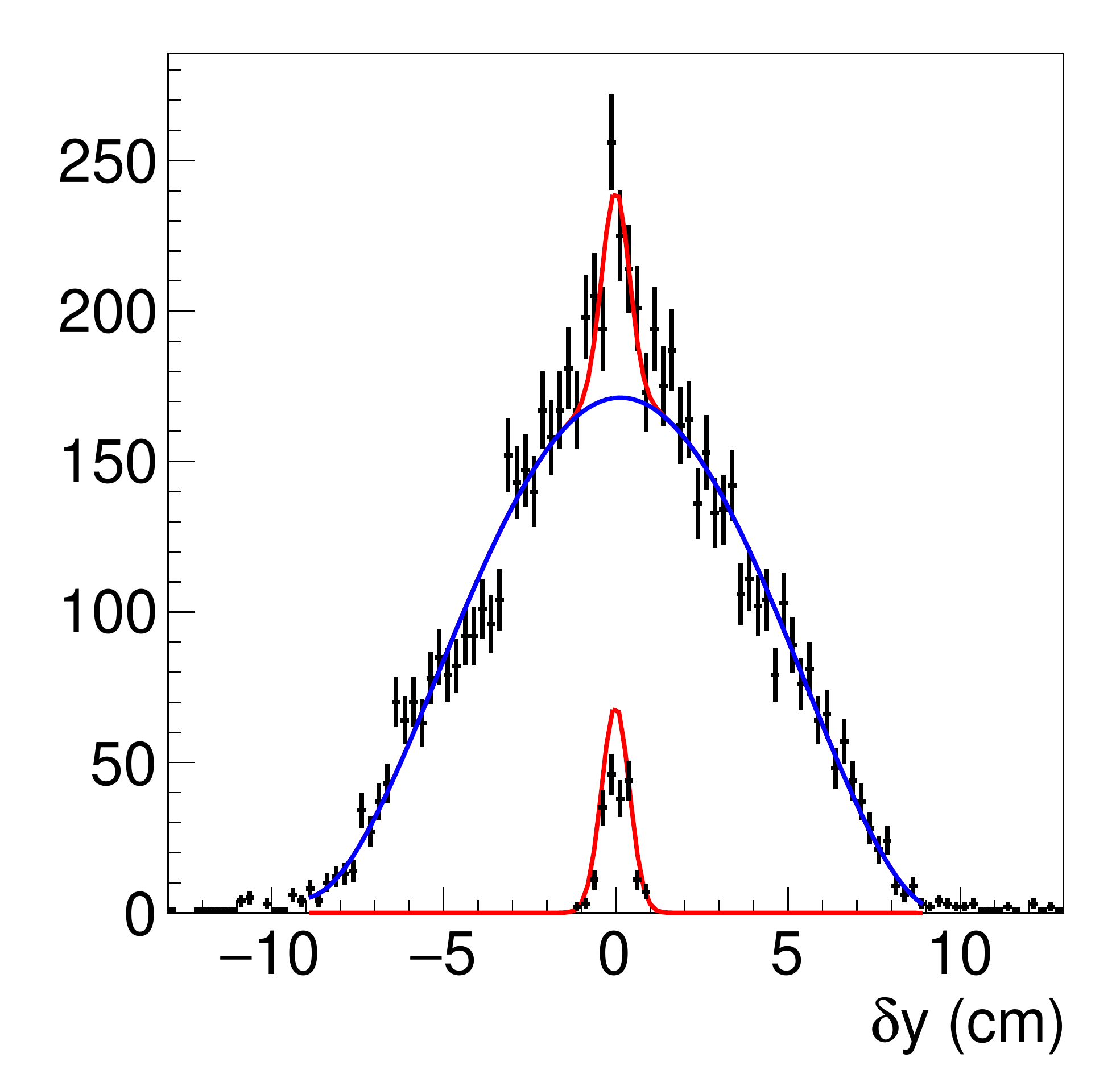}
  \end{center}
  \vspace{-5mm}
  \caption{Histogram of $\delta_y$, the difference between the
    out-of-plane measured and predicted calorimeter hit position.}
\label{fig:dy}
\end{figure}

\item \textbf{Challenges for 12 GeV}
\label{sec:12GeV}

Experience in preparing for E12-14-003~\cite{woj14}, which will extend
the measurements of the differential cross section for WACS, has
highlighted some of he challenges associated with the 12 GeV era at
Jefferson Lab, These include:
\begin{itemize} 
\item The cross section of WACS decreases much more rapidly than the
  pion photoproduction cross section, meaning a larger background.
\item The mixed electron-photon beam means radiative elastic ep
  scattering is a significant and problematic background.
\item The higher energy of the pion means the associated difference
  distribution is narrower.
\end{itemize} 

These factors, coupled with low figure-of-merit for recoil proton
polarimetry and unfavourable spin transport in the proton
spectrometers, almost certainly preclude any future measurement of the
final-state helicity observable $K_{LL}$. Efforts are therefore
focused on measurements of $A_{LL}$ with a polarized ammonia target.
While driven primarily by concerns in terms of target depolarization,
the development of a pure photon source has the added advantage of
eliminating the $ep\gamma$ background, making the data analysis much
more straightforward and causing a significant reduction in the
associated systematic uncertainties. 

Moving from liquid hydrogen to a polarized ammonia target means one
can expect more multiple scattering of the recoil proton due to the
higher density and atomic number. This then suggests that the impact
of the photon beam spot size on the target could be significant, as
even a small change in the resultant angular resolution can have a
large impact on the precision with which one can extract WACS
observables.


\item \textbf{Simulation Results}
\label{sec:sim}

In order to quantify the potential impact on the accuracy of
measurements, specifically, of $A_{LL}$, Geant4~\cite{ago03}
simulations were performed for a kinematic setting of $s =
12~\mathrm{GeV^2}$, $-t = 5~\mathrm{GeV^2}$
($\Theta_{CM}=90^{\circ}$). The experimental set-up included a 9 \%
copper radiator, the polarized ammonia target and scattering chamber,
the Super Bigbite Spectrometer (SBS) for detection of recoil protons
at $25^{\circ}$ in the lab and the Neutral Particle Spectrometer (NPS)
at $28^{\circ}$ and a distance of 3m.  Calculations of statistical
precision were performed assuming 200 hours of 8.8 GeV beam at a
current of 1 $\mathrm{\mu A}$ according to:

$$ \Delta A_{LL} = \frac{1}{\sqrt{\frac{N_{RCS}}{D}} P_p P_{\gamma}},$$

where $D=N_{RCS}/(N_{RCS}+N_{\pi^0})$ is the dilution factor due to the pion background, $P_p$ and
$P_{\gamma}$ are the values for target and beam polarization, and

$$ N_{RCS} = \frac{d\sigma}{dt} \frac{(E'_{\gamma})^2}{\pi} ( \Delta \Omega_{\gamma} f_{\gamma p})  N_{\gamma} N_p.$$

Both the solid angle acceptance factor $( \Delta \Omega_{\gamma}
f_{\gamma p})$ and incident photon flux $N_{\gamma}$ were directly
determined in the simulation.

Several different photon beam configurations were investigated, based
on varying the radiator-to-target distance and whether the beam was
collimated (2 mm diameter) or not. The results are shown for simulated
data over an incident energy range ($E_{in}$) of 5 - 6 GeV in
Fig.~\ref{fig:target1}, and for a range closer to the endpoint of 7 -
8 GeV in Fig.~\ref{fig:target2}. The linear increase in statistical uncertainty observed in both
figures as the radiator-to-target distance becomes larger is due
predominately to an increase in the pion dilution factor. This itself
is due to an increase in the photon beam spot size resulting in poorer
angular resolution. One can also clearly see the effects of
collimation as a result of reduced incident flux. The poorest
statistical uncertainty (8 m, collimated) is a factor of around 1.2
larger than the corresponding value for the \emph{standard}
configuration (a radiator-to-target distance of 73 cm, as used in the
6 GeV experiments).

\begin{figure}
  \begin{center}
    \includegraphics[width=0.45\textwidth]{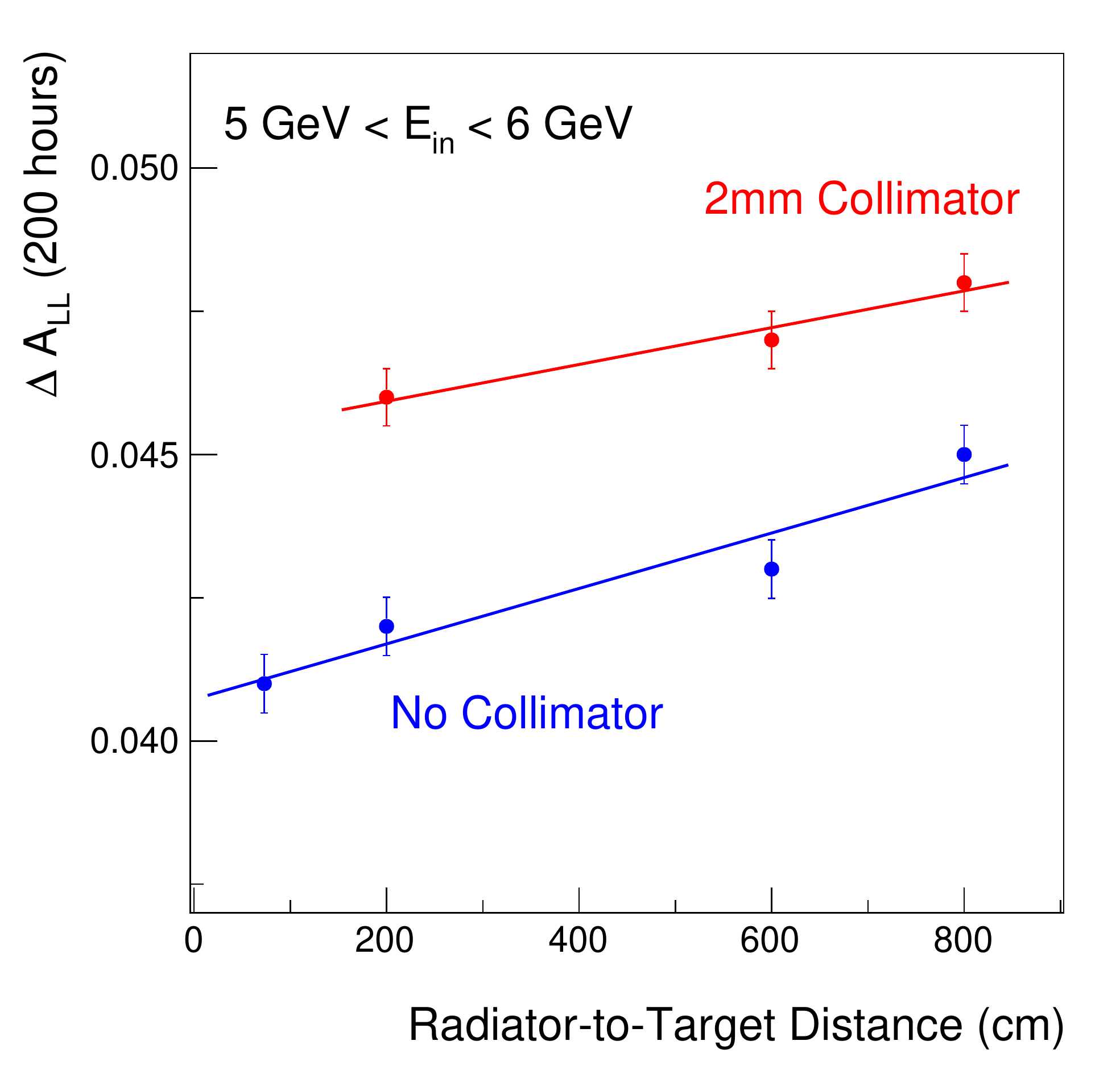}
  \end{center}
  \vspace{-5mm}
  \caption{$\Delta A_{LL}$ versus radiator-to-target distance for an
    incident energy range of 5 - 6 GeV (the blue points are without a
    collimator, while the red points are for a collimated photon beam).}
\label{fig:target1}
\end{figure}

\begin{figure}[!h]
  \begin{center}
    \includegraphics[width=0.45\textwidth]{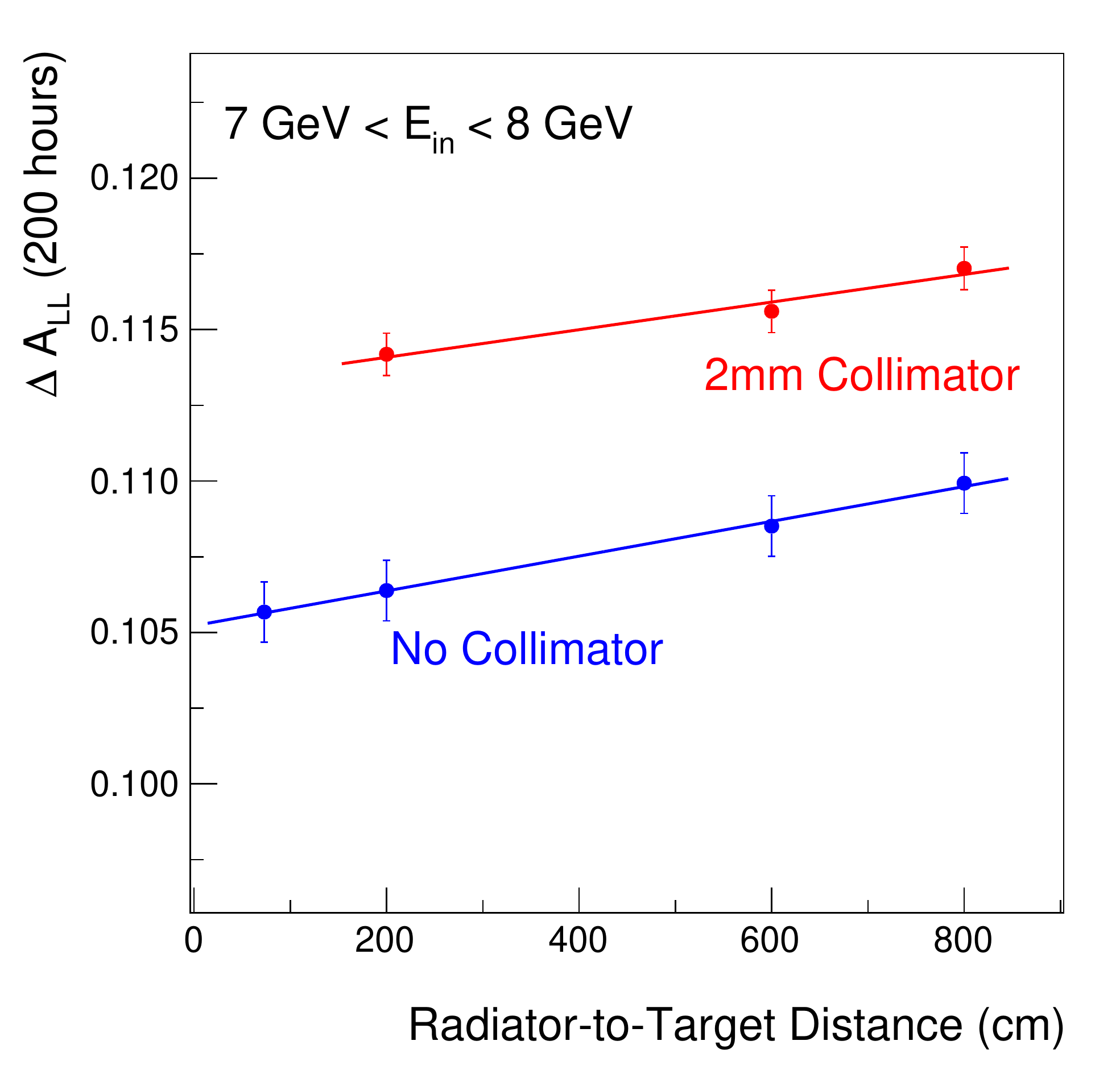}
  \end{center}
  \vspace{-5mm}
  \caption{$\Delta A_{LL}$ versus radiator-to-target distance for an
    incident energy range of 7 - 8 GeV (the blue points are without a
    collimator, while the red points are for a collimated photon beam).}
\label{fig:target2}
\end{figure}


\item \textbf{Conclusion}
\label{sec:conc}

The fact that the simulated results demonstrate an increase in beam
spot size causes a decrease in the statistical precision of $A_{LL}$
is perhaps not surprising. In fact, these results demonstrate the
impact is not as large as one might expect. Indeed, the results show
that the reduction in incident photon flux as a result of collimating
the photon beam has a more significant, but still not dramatic, impact
on the statistical uncertainty. Of course, these results do not imply
that the photon beam should not be collimated, as other factors such
as background rates in the detectors and radiation levels in the hall
must be taken into account. Also, there has been no attempt made in
the present study to optimise the collimator design or position along
the beamline.


\end{enumerate}


\newpage
\subsection{Physics Perspectives for Future K-Long Facility}
\addtocontents{toc}{\hspace{2cm}{\sl Igor~I.~Strakovsky}\par}
\setcounter{figure}{0}
\setcounter{table}{0}
\setcounter{equation}{0}
\setcounter{footnote}{0}
\halign{#\hfil&\quad#\hfil\cr
\large{Igor~I.~Strakovsky\footnote{Email:  igor@gwu.edu}}\cr
\textit{Institute for Nuclear Studies, Department of Physics}\cr
\textit{The George Washington University}\cr
\textit{Washington, D.C. 20052, U.S.A.}\cr}

\begin{abstract}
Our main interest in creating a secondary high-quality neutral 
Kaon-beam is to investigate hyperon spectroscopy. The experiment should 
measure both differential cross sections and self-analyzed polarizations 
of the produced $\Lambda$-, $\Sigma$-, and $\Xi$-hyperons using the 
GlueX detector at the Jefferson Lab Hall~D and both proton and "neutron" 
targets. New data will greatly constrain partial-wave analysis and 
reduce model-dependent uncertainties in the extraction of strange 
resonance properties, providing a new benchmark for comparisons with 
QCD-inspired models and LQCD calculations.
\end{abstract}

\begin{enumerate}
\item \textbf{Introduction}

Three light quarks can be arranged in 6 baryonic families,
$N^\ast$, $\Delta^\ast$, $\Lambda^\ast$, $\Sigma^\ast$,
$\Xi^\ast$, and $\Omega^\ast$.  The number of members in a family
that can exist is not arbitrary~\cite{Nefkens1997}.  If
SU(3)$_F$ symmetry of QCD is controlling, then for the octet:
$N^\ast$, $\Lambda^\ast$, and $\Sigma^\ast$, and for the
decuplet: $\Delta^\ast$, $\Sigma^\ast$, $\Xi^\ast$, and
$\Omega^\ast$.  The number of ``experimentally" identified resonances
of each baryon family in the PDG2016 summary tables is 17 $N^\ast$,
24 $\Delta^\ast$, 14 $\Lambda^\ast$, 12 $\Sigma^\ast$, 7
$\Xi^\ast$, and 2 $\Omega^\ast$~\cite{PDG2016}.  Constituent QMs, 
for instance, predict existence of no less than 64 $N^\ast$ and 
22 $\Delta^\ast$ states with mass less than 3~GeV.  The seriousness
of "missing-states" problem~\cite{Koniuk1980} is obvious from
these numbers. To complete SU(3)$_F$ multiplets, one needs no
less than 17 $\Lambda^\ast$, 41 $\Sigma^\ast$, 41 $\Xi^\ast$,
and 24 $\Omega^\ast$.

Our current "experimental" knowledge on the $\Lambda^\ast$,
$\Sigma^\ast$, $\Xi^\ast$, and $\Omega^\ast$ resonances is
far worse than our knowledge of the $N^\ast$ and $\Delta^\ast$
ones; though they are equally fundamental. The Breit-Wigner 
masses and widths of the lowest $\Lambda$ and $\Sigma$ baryons 
were determined mainly with charged Kaon-beam experiments in 
the 1970s~\cite{PDG2016}.  Pole position in complex energy 
plane for hyperons has began to be studied only recently, first 
of all for $\Lambda(1520)\frac{3}{2}^-$~\cite{Qiang2010}.  
Clearly, complete understanding of three-quark bound states 
requires to learn about baryon resonances in ``strange" sector.  
Specifically, the properties of multi-strange baryons are poorly 
known. For instance the {\it Review of Particle Physics} lists 
only two states with BR to $K\Xi$, namely, $\Lambda(2100)\frac{7}
{2}^-$ (BR $<$ 3\%) and $\Sigma(2030)\frac{7}{2}^+$ (BR $<$ 
2\%)~\cite{PDG2016}.  

There are two particles in the reactions $K_Lp\rightarrow\pi Y$
and $KY$ that can carry polarization: the target and recoil
nucleon/hyperon. Hence, there are two possible double-polarization 
experiments: target/recoil. While a formally complete experiment 
requires the measurement, at each energy and angle, of at least 
three independent observables, the current database for 
$K_Lp\rightarrow\pi Y$ and $KY$ is populated mainly by unpolarized 
cross sections~\cite{Brem}. Meanwhile, the quality of avilable P
measurements do not have a sensitivity to the PWA 
fits~\cite{Mark2016}.

The experiments using unpolarized LD$_2$ (to get "neutron"
data) will open up a new avenue to the complete experiment. Note 
that the "neutron" data are critical to determine parameters of 
neutral $\Lambda^\ast$s and $\Sigma^\ast$s hyperons which were 
considered recently~\cite{Zou2016}.

Overall, limited number of $K_L$ induced measurements on proton 
target (1961 -- 1982) 2426~d$\sigma$/d$\Omega$, 348~$\sigma^{tot}$, 
and 115~P observables do not allow today to feel comfortable with 
Hyperon Spectroscopy results~\cite{Brem}.

The JLab12 energy upgrade, with the new Hall~D, is an ideal
tool for extensive studies of non-strange and, specifically,
hyperon resonances~\cite{Curtis2016}.  However due to a very low 
photoproduction cross sections, experimental studies are very 
limited. Our plan is evolving to take advantage of the existing 
high quality photon beam line and experimental area in the Hall~D 
complex at Jefferson Lab to deliver a beam of $K_L$ particles 
onto liquid hydrogen or deuterium cryotargets within the GlueX 
detector.  The recently constructed GlueX detector in Hall~D is 
a large acceptance spectrometer with good coverage for both 
charged and neutral particles that can be adapted to this purpose.
The experimental setup is well suited to measure both differential 
cross sections and self-analyzed polarizations of the 
produced $\Lambda$-, $\Sigma$-, and $\Xi$-hyperons using the GlueX 
detector at the Jefferson Lab Hall~D and both proton and "neutron" 
targets.

At the first stage, E$_e$ = 12~GeV electrons produced at the CEBAF 
will scatter in a radiator generating intensive beam of 
bremsstrahlung photons (we will not need in the Hall~D Broadband 
Tagging Hodoscope).  At the second stage, bremsstrahlung photons, 
created by electrons, hit the Be-target and produce neutral Kaons 
along with neutrons and charged particles.  The charged particles 
are deflected by the magnetic field of a magnet installed after the
Be-target. The neutron yield is negligible on the GlueX target. 
Finaly, $K_L$s will reach the LH$_2$ or LD$_2$ cryogenic target 
within GlueX settings. The measurements will span c.m. $\cos\theta$ 
from -0.95 to 0.95 in c.m. range above W = 1490~MeV and up to 4000~MeV.

We estimated the flux of $K_L$ beam on the GlueX LH$_2$
target is about $3\times 10^4~K_L/s$, almost comparable
to charged Kaon rates obtained at AGS and elsewhere in
the past and expected for J-PARC~\cite{Noumi2016}. Momenta
of neutral Kaons at JLab will be measured applying the
time-of-flight technique using a time structure of 64~ns.
The count rate estimates carried out assuming 100~days of
data taking are presented in Fig.~\ref{fig:fig5}(bottom row)
which can be compared with available buble chamber measurements 
Fig.~\ref{fig:fig5}(top row).
\begin{figure}[ht]
\centerline{
\includegraphics[height=0.42\textwidth, angle=90]{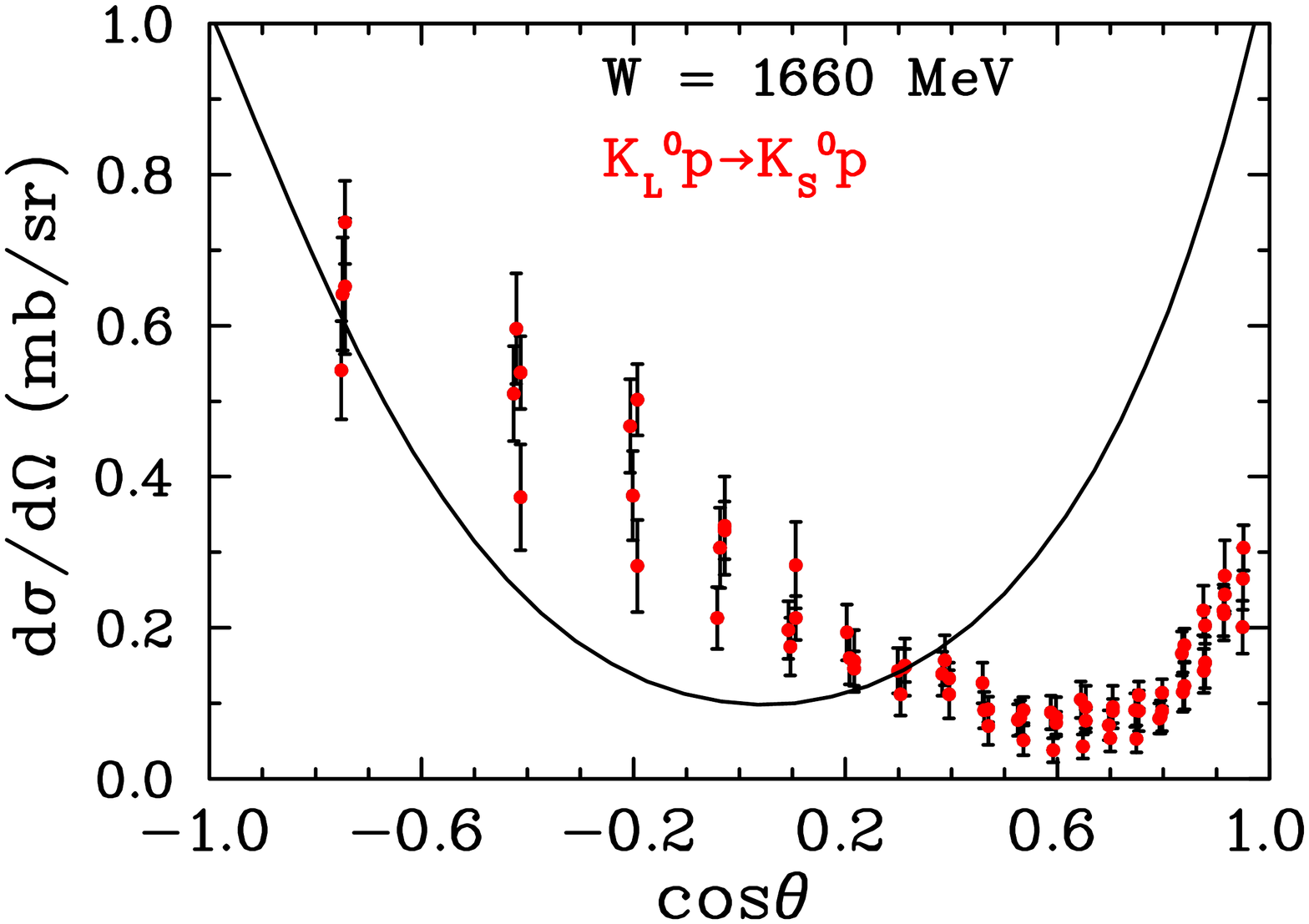}\\
\includegraphics[height=0.42\textwidth, angle=90]{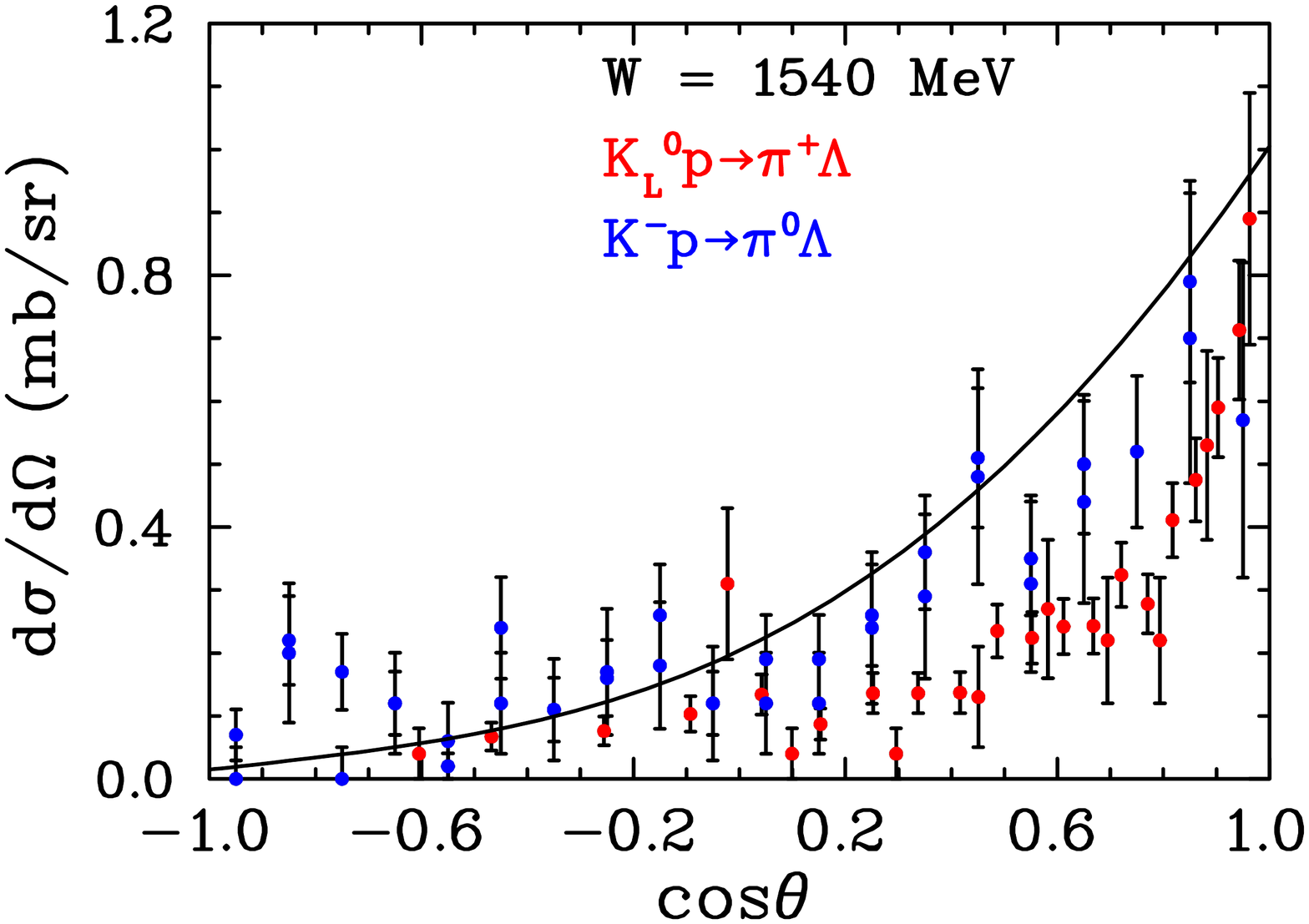}}
\centerline{
\includegraphics[height=0.31\textwidth, angle=0]{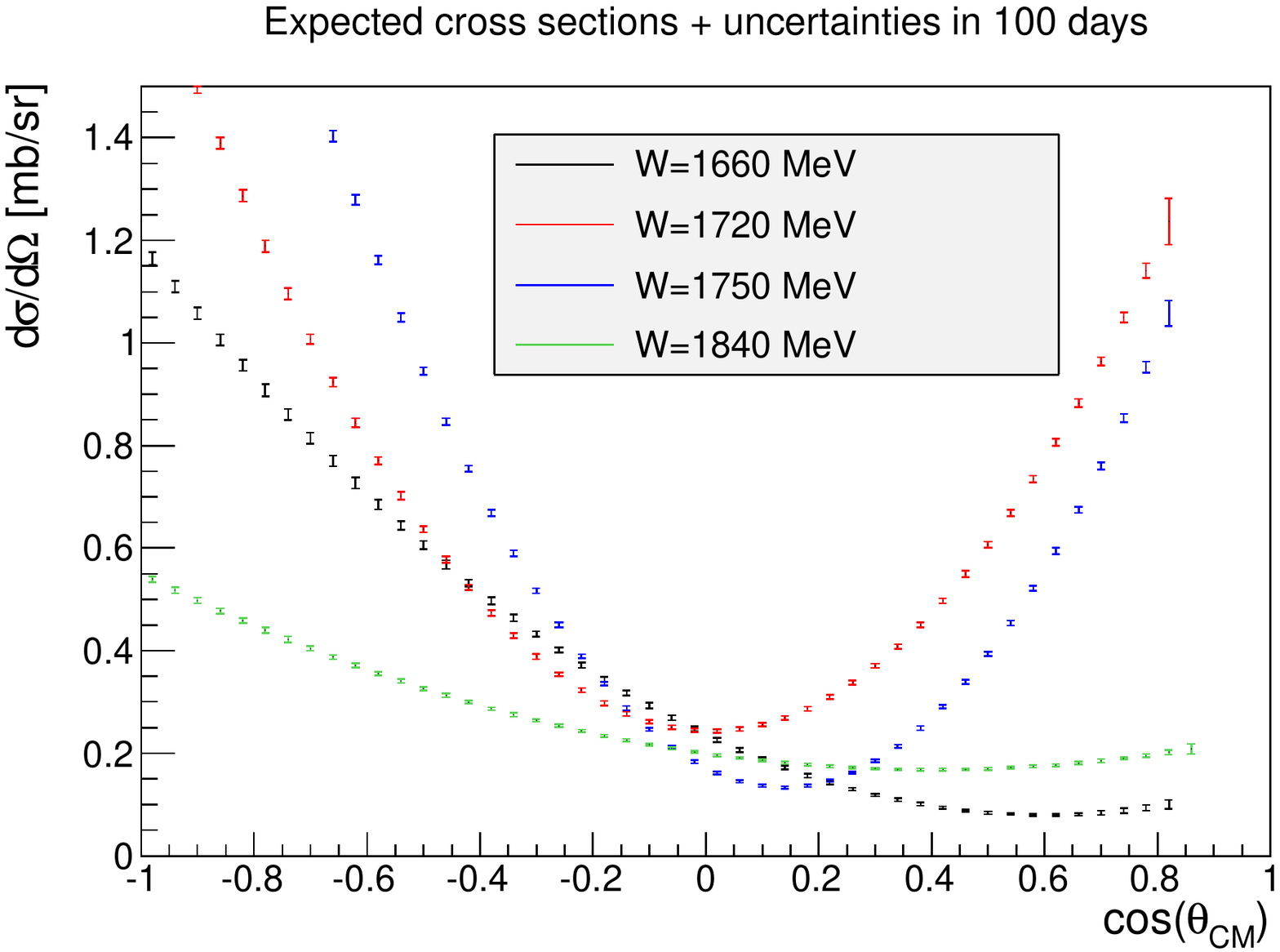}\\
\includegraphics[height=0.31\textwidth, angle=0]{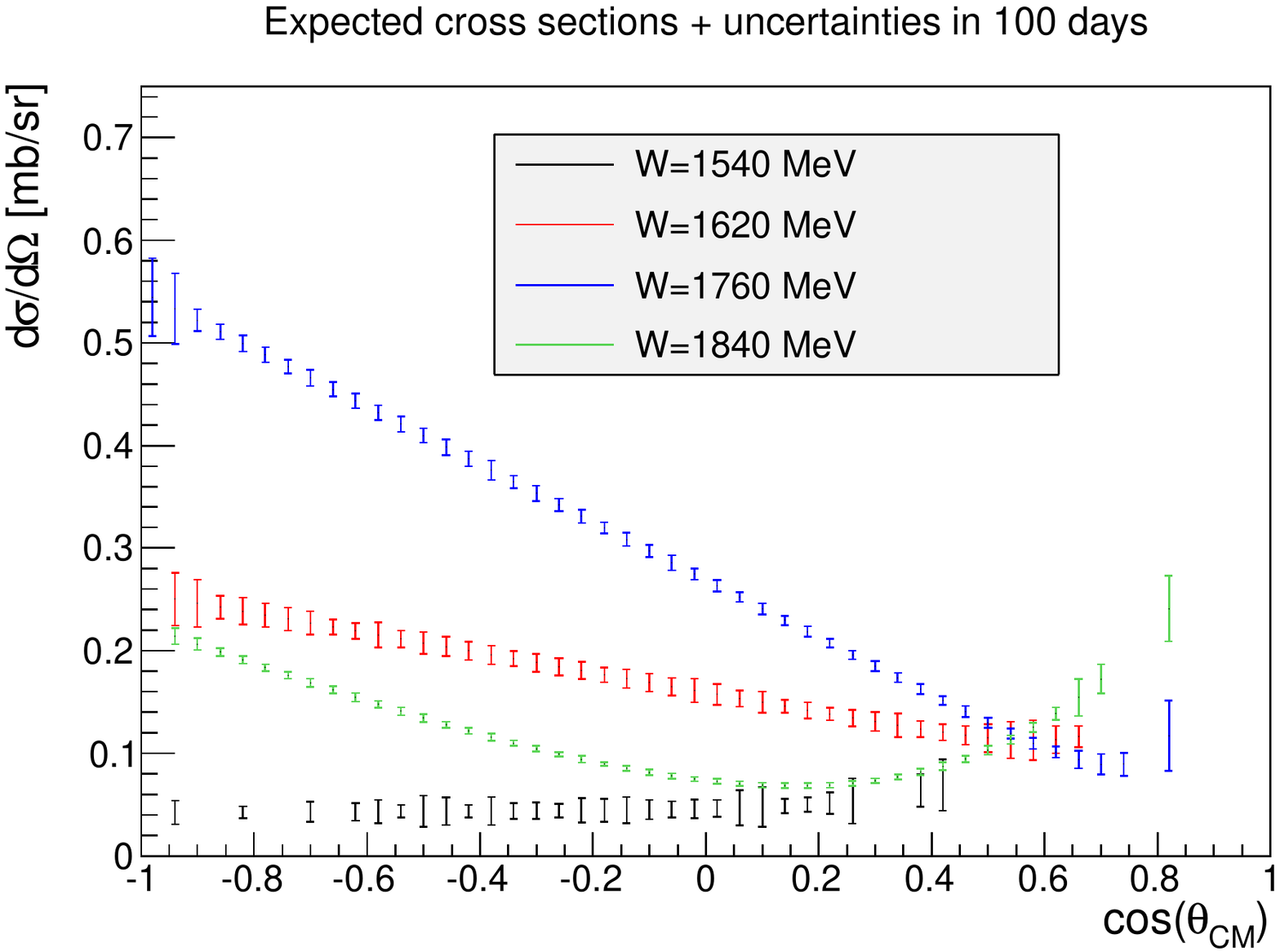}}

\caption{Samples of cross section for $K_Lp\rightarrow pK_S$ (left)
        and $K_Lp\rightarrow\pi^+\Lambda$ (right).
        Old bubble chamber data and curves are from the recent
        PWA~\protect\cite{Zhang2013} (top row) and expected
        KL measurements using full Monte Carlo calculations
	(bottom row). \label{fig:fig5}}
\end{figure}

There is an advantage factor for $K_Lp$ vs. $K^-p$ experiment.
The mean lifetime of the $K_L$ is 51.16~ns ($c\tau = 15.3$~m)
whereas the mean lifetime of the $K^-$ is 12.38~ns ($c\tau =
3.7$~m)~\cite{PDG2016}.  For this reason, it is much easier to
perform measurements of $K_Lp$ scattering at low beam energies
compared with $K^-p$ scattering~\cite{Noumi2016}.

New data will greatly constrain partial-wave analysis and
reduce model-dependent uncertainties in the extraction of
hyperon properties, providing a new benchmark for
comparisons with QCD-inspired models and LQCD calculations.
We plan to do a coupled-channel PWA (following H\"ohler's
prescription~\cite{Hoehler1983}) with new GlueX data in
combination with available and new J-PARC $K^-p$ measurements.  
Then the best fit will allow to determine data driven (model 
independent) partial-wave amplitudes and associated resonance 
parameters as the SAID group does, for instance, for analysis 
of $\pi$N-elastic, charge-exchange, and $\pi^-p\rightarrow\eta 
n$ data~\cite{Arndt2006}.  With the new GlueX data, the 
quantitative significance of resonance signals can be 
determined.  Additionally, new PWA with new GlueX data will 
allow to look for "missing" hyperons via looking for new poles 
in complex plane positions.  

Precise new data (both differential cross section and recoil 
polarization of hyperons) for $K_Lp$ scattering with good 
kinematic coverage could significantly improve our knowledge on
$\Lambda^\ast$, $\Sigma^\ast$, $\Xi^\ast$, and $\Omega^\ast$ 
resonances.  Clearly, complete 
understanding of three-quark bound states requires to learn 
more about baryon resonances in "strange sector".  Polarization 
data are very important to be measured in addition to 
differential cross sections to help remove ambiguities in PWAs. 

Measurements of "missing" hyperon states with their spin-parity  
assignments along with the "missing" non-strange baryons will 
provide very important ingredients to test QM and LQCD 
predictions thereby improving our understanding of QCD in a 
non-perturbative regime.\\

\item \textbf{Acknowledgments}
I thank Moskov Amaryan, Yakov Azimov, William Briscoe, Eugene 
Chudakov, Pavel Degtyarenko, Michael D\"oring, Alexander Laptev, 
Ilya Larin, Maxim Mai, Mark Manley, James Ritman, and, Simon 
Taylor for comments on the feasibility of future measurements. 
This work is supported, in part, by the U.S.~Department of 
Energy, Office of Science, Office of Nuclear Physics, under 
Award Number DE--SC0016583.

\end{enumerate}


\newpage
\subsection{Structure of the antikaon-nucleon scattering amplitude}
\addtocontents{toc}{\hspace{2cm}{\sl M.~Mai}\par}
\setcounter{figure}{0}
\setcounter{table}{0}
\setcounter{equation}{0}
\setcounter{footnote}{0}
\halign{#\hfil&\quad#\hfil\cr
\large{Maxim Mai \footnote{Email:  maximmai@gwu.edu}}\cr
\textit{Institute for Nuclear Studies and Department of Physics}\cr
\textit{The George Washington University}\cr
\textit{Washington, DC 20052, U.S.A.}\cr}

\begin{abstract}
In this talk we discuss the modern approaches of the antikaon-nucleon scattering based on unitarization of the interaction potential derived from Chiral Perturbation Theory. Such approaches rely on very old experimental data, when fitting their free parameters. Thus, ambiguities arise in the form of obtained scattering amplitudes. We demonstrate the occurance of these ambiguities on the example of one specific framework and discuss several possibilities to overcome these.
\end{abstract}

\begin{enumerate}
\item \textbf{Introduction}

Antikaon-nucleon scattering is one of the most discussed reactions among the meson-baryon interaction channels from the corresponding ground state octets. At energies below the chiral symmetry breaking scale, Chiral Perturbation Theory (ChPT) allows one to investigate the meson-baryon scattering amplitudes systematically, see Ref.~\cite{Bernard:2007zu} for a recent review on baryon ChPT. The full calculation of the scattering length up to next-to-next-to-leading chiral order, see Ref.~\cite{Mai:2009ce}, yields for the isoscalar channel of the  antikaon-nucleon scattering (in fm)
\begin{align*}
a_{\bar KN}^{I=0}=(+0.53)_{\rm LO}+(+0.97)_{\rm NLO}+(-0.40+0.22i)_{\rm NNLO}=+1.11+0.22i\,.
\end{align*}
The convergence of the series is rather slow, which is presumably due to the large kaon mass as well as large separation of the coupled channels thresholds. Further, the net result disagrees with the experimental one ($a_{\bar KN}^{I=0}\approx-0.53+0.77i$~fm) even in the sign. This result is derived from from the measurement of the energy shift and width of kaonic hydrogen in the SIDDHARTA experiment at DA$\Phi$NE \cite{Bazzi:2011zj}, using the Deser-type formula from Ref.~\cite{Meissner:2004jr}. The main reason for this behavior is the presence of a sub-threshold resonance, the so-called $\Lambda(1405)$ in this channel, which was already indicated in the early studies~\cite{Dalitz:1959dq}. Therefore, a perturbative treatment inevitably breaks down and non-perturbative techniques are required.

The chiral unitary approach (UChPT) is considered to be the best non-perturbative framework to address the SU(3) dynamics of the antikaon-nucleon system at low-energies. Many studies have been performed in the last two decades using different versions of this framework, see, e.g., Refs.~\cite{Borasoy:2006sr,Oset:1997it,Jido:2003cb,Hyodo:2011ur,Kaiser:1995eg,Mai:2014xna,Mai:2012dt,Oller:2000fj,Cieply:2011nq} as well as Ref.~\cite{Hyodo:2011ur} for an extensive review. The main purpose of this work is to study this ambiguity as well as to suggest ways to reduce it. To perform this study in a systematic manner, we will restrict ourselves here to one single, but the most general framework derived in Refs.~\cite{Bruns:2010sv,Mai:2014xna,Mai:2012dt}. We will show that at least several solutions for the antikaon-nucleon scattering amplitude agree with the experimental scattering data. However, including data, e.g, from photoproduction experiment at CLAS~\cite{Moriya:2014kpv}, some of these solutions can be disregarded as unphysical. Further, using synthetic data we discuss the possible impact of the new measurements of cross sections, which might become available in the proposed $K_{\rm long}$-beam experiment at Jefferson Lab~\cite{Amaryan:2015swp}.

\item \textbf{Analysis of scattering data}

For the analysis of the presently available scattering data we rely on the amplitude constructed and described in detail in Refs.~\cite{Mai:2014xna,Mai:2012dt,Bruns:2010sv}, to which we refer the reader for the conceptual details. We start from the chiral Lagrangian of the leading (LO) and next-to-leading (NLO) chiral order, see Refs.~\cite{Krause:1990xc,Frink:2004ic}. For the reasons explained in Refs.~\cite{Mai:2014xna,Mai:2012dt,Bruns:2010sv}, the $s$- and $u$-channel one-baryon exchange diagrams are neglected, which reduces the form of the chiral potential to
\begin{align}\label{eqn:potential}
 V(\slashed{q}_2, \slashed{q}_1; p)=&A_{WT}(\slashed{q_1}+\slashed{q_2})
 +A_{14}(q_1\cdot q_2)+A_{57}[\slashed{q_1},\slashed{q_2}]
 +A_{M} +A_{811}\Big(\slashed{q_2}(q_1\cdot p)+\slashed{q_1}(q_2\cdot p)\Big)\,,
\end{align} 
where the incoming and outgoing meson four-momenta are denoted by $q_1$ and $q_2$, respectively. The overall four-momentum of the meson-baryon system is denoted by $p$. The symbols $A_{WT}$, $A_{14}$, $A_{57}$, $A_{M}$ and $A_{811}$ denote the 10-dimensional matrices which encode the coupling strengths between all 10 channels of the meson-baryon system of strangeness $S=-1$, i.e. $\{K^-p$, $\bar K^0 n$, $\pi^0\Lambda$, $\pi^0\Sigma^0$, $\pi^+\Sigma^-$, $\pi^-\Sigma^+$, $\eta\Lambda$, $\eta \Sigma^0$, $K^+\Xi^-$, $K^0\Xi^0\}$. These matrices  depend on the meson decay constants, the baryon mass in the chiral limit, the quark masses as well as 14 low-energy constants (LECs) as specified in the original publication~\cite{Mai:2014xna}.

The above potential is used as a driving term of the coupled-channel Bethe-Salpeter equation (BSE), for the meson-baryon scattering amplitude $T$
\begin{align}\label{eqn:BSE}
T(\slashed{q}_2, \slashed{q}_1; p)=V(\slashed{q}_2, \slashed{q}_1;p)   +i\int\frac{d^d l}{(2\pi)^d}V(\slashed{q}_2, \slashed{l}; p)   S(\slashed{p}-\slashed{l})\Delta(l)T(\slashed{l}, \slashed{q}_1; p)\,,
\end{align}
where $S$ and $\Delta$ represent the baryon (of mass $m$) and the meson (of mass $M$) propagator, respectively, and are given by $iS(\slashed{p}) = {i}/({\slashed{p}-m+i\epsilon})$ and $i\Delta(k) ={i}/({k^2-M^2+i\epsilon})$. Moreover, $T$, $V$, $S$ and $\Delta$ in the last expression are matrices in the channel space. The loop diagrams appearing above are treated using dimensional regularization and applying the usual $\overline{\rm MS}$ subtraction scheme in the spirit of Ref.~\cite{Bruns:2010sv}. Note that the modified loop integrals are still scale-dependent. This scale $\mu$ reflects the influence of the higher-order terms not included in our potential. It is used as a fit parameter of our approach. To be precise, we have 6 such parameters, neglecting isospin breaking effects. The above equation can be solved analytically, if the kernel contains contact terms only as shown in Ref.~\cite{Bruns:2010sv}. Using this solution for the strangeness $S=-1$ system, we have shown in Ref.~\cite{Mai:2012dt} that once the full off-shell amplitude is constructed, one can easily reduce it to the on-shell solution, i.e. setting all tadpole integrals to zero. It appears that the double pole structure of the $\Lambda(1405)$ is preserved by this reduction and that the position of the two poles are changing only by about $20$~MeV in the imaginary part. On the other hand, the use of the on-shell approximation of the Eq.~\eqref{eqn:BSE} reduces the computational time roughly by a factor of 30. Therefore, since we wish to explore the parameter space in greater detail, it seems to be safe and also quite meaningful to start from the solution of the BSE~\eqref{eqn:BSE} with the chiral potential \eqref{eqn:potential} on the mass-shell. When required, the off-shell solution can then be obtained, gradually turning the off-shell terms on again.

The free parameters of the present model, the low-energy constants and the regularization scales $\mu$ are adjusted  to reproduce all known experimental data in the meson-baryon sector. These are the cross sections for the processes $K^-p\to K^-p$, $K^0n$, $\pi^0\Lambda$, $\pi^+\Sigma^-$, $\pi^0\Sigma^0$, $\pi^-\Sigma^+$~\cite{Ciborowski:1982et,Humphrey:1962zz,Sakitt:1965kh,Watson:1963zz} for laboratory momentum $P_{\rm lab}<300$ MeV. Electromagnetic effects are not included in the analysis and are assumed to be negligible at the measured values of $P_{\rm lab}$. Additionally, we consider the following threshold decay ratios from Refs.~\cite{Tovee:1971ga,Nowak:1978au} as well as the energy shift and width of kaonic hydrogen in the 1s state, i.e. $\Delta E -i\Gamma/2=(283\pm42)-i(271\pm55)$~eV from the SIDDHARTA experiment at DA$\Phi$NE \cite{Bazzi:2011zj}. The latter two values are related to the $K^-p$ scattering length via the modified Deser-type formula \cite{Meissner:2004jr}. 

Assuming naturalness of the free parameters of the model we obtain eight best solutions in fits to the aforementioned data, see second row of the Tab.~\ref{tab:photo}. The data are described equally well by all eight solutions, showing, however, different functional behavior of the cross sections as a function of $P_{\rm lab}$. These differences are even more pronounced for the scattering amplitude $f_{0+}$, which is fixed model independently only in the ${K^-p}$ channel at the threshold by the scattering length $a_{K^-p}$. Similar observation was made in the comparison of this approach with other most recently used UChPT models in Ref.~\cite{Cieply:2016jby}. 

When continued analytically to the complex energy plane, all eight solutions confirm the double pole structure of the $\Lambda(1405)$ on the second Reimann Sheet. The scattering amplitude is restricted around the $\bar K N$ threshold by the SIDDHARTA measurement quite strongly. Therefore, in the complex energy plane we observe a very stable behavior of the amplitude at this energy, i.e. the position of the narrow pole agrees among all solutions within the $1\sigma$ parameter errors. This is in line with the findings of other groups~\cite{Ikeda:2012au,Borasoy:2006sr,Guo:2012vv}, i.e. one observes stability of the position of the narrow pole. The position of the second pole is, on the other hand, less restricted. To be more precise, for the real part we find three clusters of these poles: 
around the $\pi\Sigma$ threshold, around the $\bar K N$ threshold as well as around $1470$~MeV. For several solutions there is some agreement in the positions of the second pole between the present analysis and the one of Ref.~\cite{Guo:2012vv} and of our previous work \cite{Mai:2012dt}. However, as the experimental data is described similarly well by all fit solutions, one can not reject any of them, which represents the systematic uncertainty of this approach.

\item \textbf{Reduction of {\bf the} model ambiguities}

\renewcommand{\baselinestretch}{1.25}
\begin{table*}[t]
\begin{center}
\begin{tabular}{|c|cccccccc|}
\hline
~~~~~~~~Fit \#~~~~~~~~ 
&~~~~1~~~~
&~~~~2~~~~
&~~~~3~~~~
&~~~~4~~~~
&~~~~5~~~~
&~~~~6~~~~
&~~~~7~~~~
&~~~~8~~~~\\
\hline
~~$\chi_{\rm d.o.f.}^2$ (hadronic data)~~ &1.35 &1.14&0.99 &0.96 &1.06 &1.02
&1.15 &0.90\\
\hline
$\chi_{\rm p.p.}^2$   (CLAS data)~~~~~ &3.18&1.94&2.56&1.77&1.90&6.11&2.93&3.14\\
\hline
\end{tabular} 
\caption{Fit quality for the hadronic~\cite{Ciborowski:1982et,Humphrey:1962zz,Sakitt:1965kh,Watson:1963zz} and photoproduction data~\cite{Moriya:2014kpv}.
\label{tab:photo}}
\end{center}
\end{table*}

In Ref.~\cite{Cieply:2016jby} a direct quantitative comparison of the present and similar other chiral unitary approaches has been made. As demonstrated there, the ambiguity of the scattering amplitudes pointed out above is similar to the one between and among other most recent chiral unitary approaches on antikaon-nucleon scattering, which rely on the same set of experimental data. Therefore, it is instructive to use the present approach to test ways to reduce this systematic uncertainty. In the following we will discuss three of such possibilities.

~\\ \noindent
\textbf{1) Photoproduction data}
Very sophisticated measurements of the reaction $\gamma p\to K^+\Sigma \pi$ were performed by the CLAS collaboration at JLab, see Ref.~\cite{Moriya:2013eb}. There, the invariant mass distribution of all three $\pi\Sigma$ channels was determined in a broad energy range and with high resolution. Finally, from these data the spin-parity analysis of the $\Lambda(1405)$ was performed in Ref.~\cite{Moriya:2014kpv}. 

To make use of these high quality data we have assumed in Ref.~\cite{Mai:2014xna} the simplest ansatz for the photoproduction amplitude similar to the one of Ref.~\cite{Roca:2013av}. Specifically, we have parameterized the two-meson production mechanism by unknown, energy-dependent coupling constants $C$, using the hadronic amplitudes from the last chapter as final state interaction. Clearly, a more sophisticated ansatz is required to address scattering and photoproduction data simultaneously, while fulfilling in the same time the gauge invariance. Such an approach can be developed along the techniques used for the analysis of the single meson photoproduction, see Refs.~\cite{Borasoy:2007ku,Mai:2012wy}. The question we wish to address here is, however, different. Namely, whether all obtained hadronic solutions allow for a good description of the photoproduction data using such a flexible ansatz. Thus, without altering the parameters of the hadronic part (8 solutions) in the photoproduction amplitude, we fit only the unknown constants $C$ to the CLAS data in all three measured final states ($\pi^+\Sigma^-$, $\pi^0 \Sigma^0$, $\pi^-\Sigma^+$) and for all 9 measured total energy values. 

The resulting values of $\chi^2$ per data point for these fits are collected in the third row of the Table~\ref{tab:photo}, whereas for the further details and error analysis we again refer to the original publication~\cite{Mai:2014xna}. The results in Table~\ref{tab:photo} show that even within such a flexible ansatz the solutions~\#1, \#3, \#6, \#7 and \#8 of the eight hadronic solutions do not allow for a decent description of the high-quality CLAS data. Consequently, only three of originally eight solutions can be considered as physical with respect to these photoproduction data, which indeed reduces the ambiguity of the antikaon-nucleon scattering amplitude substantially.

~\\ \noindent
\textbf{2) Kaonic deuterium}
The $K^-p$ scattering amplitude at the threshold is fixed very well by the strong energy shift and width of the kaonic hydrogen, measured in the SIDDHARTA experiment~\cite{Bazzi:2011zj}. However, this does not fix the full $\bar KN$ scattering amplitude at the threshold, which has two complex valued components, i.e. for isospin $I=0$ and $I=1$. To fix both components one requires another independent measurement, such as, e.g, the energy shift and width of the kaonic deuterium. Such a measurement is proposed at LNF~\cite{LNF} and J-PARC~\cite{JPARK}. Ultimately, this quantity can be related to the antikaon-deuteron scattering length by the well known Deser-type relations (see, e.g., Refs.~\cite{Deser:1954vq,Meissner:2004jr,Meissner:2006gx,physrep}) and then to the antikaon-nucleon scattering length, using an effective field theory framework, see, e.g, Ref.~\cite{Mai:2014uma}.

~\\ \noindent
\textbf{3) Additional scattering data}
The scattering data from Refs.~\cite{Ciborowski:1982et,Humphrey:1962zz,Sakitt:1965kh,Watson:1963zz} used to fix the $\bar KN$ scattering amplitude as described in the last section stem from very old bubble chamber experiments. From the theoretical point of view, improvement of these data would be the simplest way to reduce the ambiguity of the theoretical predictions. The proposed measurement of the two-body interaction of $K_{\rm long}$-beam and the proton target at JLab~\cite{Amaryan:2015swp} can potentially lead to such an improvement of the data. In order to quantify this statement we test the already obtained solutions of the hadronic model of the last section with respect to a set of new \textit{synthetic} data on total cross sections in the same channels as before. For the synthetic data we use our best solution (\#4) in the momentum interval $P_{\rm lab}=100-300$~MeV, assuming the energy binning to be fixed but randomizing the data by Gaussian distribution with a standard deviation of $\Delta\sigma$ for the charged and $2\Delta\sigma$ for the neutral final state channels. The latter is assumed to account for the fact that neutral channels are usually more intricate to measure.

We have tested different scenarios - considering different energy binning ($\Delta P$), measurement accuracy ($\Delta\sigma$) and whether the new synthetic data complements or replaces the old data. Without further fitting, we have compared the new $\chi^2_{\rm d.o.f.}$ of all solutions, obtained in the previous section. We found that at least four of the obtained eight solutions are not compatible with the updated data for $\Delta \sigma \le 5$~MeV and $\Delta P\sim 10$~MeV. This procedure appears to be more sensitive to the measurement accuracy than on the energy binning - for $\Delta\sigma\ge10$~MeV none of the solutions could be sorted out for any of chosen values of $\Delta P$. A complete replacement of the old by the new (synthetic) data does not change our findings qualitatively, but increases the differences between new $\chi^2_{\rm d.o.f.}$ values slightly. In summary, this preliminary and simplistic analysis underlines the importance of the re-measurement of the cross section data on $\bar K N$ scattering in a modern experimental setup, such as the one proposed in Ref.~\cite{Amaryan:2015swp}.

\item \textbf{Acknowledgments}
The speaker is grateful to the organizers of the workshop for the invitation. The results presented in this talk are based on several studies and multiple discussions over the last years with Ulf-G. Mei{\ss}ner, A. Ciepl\'y, P. Bruns, M. D\"oring and D. Sadasivan. The speaker is grateful for the financial support of the German Research Foundation (DFG) under the fellowship MA 7156/1-1.

\end{enumerate}


\newpage
\subsection{The Role of Hadron Resonances in Hot Hadronic Matter}
\addtocontents{toc}{\hspace{2cm}{\sl J.L.~Goity}\par}
\setcounter{figure}{0}
\setcounter{table}{0}
\setcounter{footnote}{0}
\setcounter{equation}{0}
\halign{#\hfil&\quad#\hfil\cr
\large{Jos\'e~L.~Goity}\cr
\textit{Department of Physics}\cr
\textit{Hampton University}\cr
\textit{Hampton, VA 23668, U.S.A. \&}\cr
\textit{Thomas Jefferson National Accelerator Facility}\cr
\textit{Newport News, VA 23606, U.S.A.}\cr}

\begin{abstract}
Hadron resonances can play a significant role in hot hadronic matter. 
Of particular interest for this workshop are the contributions of 
hyperon resonances. The question about how to quantify the effects 
of resonances is here addressed. In the framework of the hadron 
resonance gas, the chemically equilibrated case, relevant in the 
context of lattice QCD calculations, and the chemically frozen case 
relevant in heavy ion collisions are discussed.
\end{abstract}

\begin{enumerate}
\item \textbf{Introduction}

Lattice QCD (LQCD) and high energy heavy collisions (RHICs) give access 
to QCD thermodynamics in the limit of low or vanishing conserved charges 
(Baryon number B, strangeness S or electric charge Q). While LQCD 
addresses the case of chemically equilibrated hot matter, which 
corresponds ot the early universe, HICs produces a fireball which 
expands too fast for chemical equilibrium to be maintained giving rise 
to a hot hadronic system which at kinetic freeze out is well off chemical 
equilibrium. At temperatures between 0.15 -- 0.17~GeV a cross over 
transition occurs from a quark-gluon to a hadronic phase. The rigorous 
description of the hadronic phase is in principle possible with a full 
knowledge of the S-matrix~\cite{Dashenj}. Absent that knowledge, one needs 
to consider models. The simplest model, known as the hadron resonance gas 
(HRG) turns out to provide a remarkably good  description of thermodynamic 
observables.
The HRG  is to a first approximation an ideal gas of hadrons, consisting 
of mesons and baryons and their resonances. The HRG gas is then determined
simply by the hadron spectrum. The hadron spectrum is however
incompletely known, in particular for baryons, and thus one question is
how important the role of such "missing" states may be in the HRG; this
issue is the main focus of this note. Several indications of missing
states exist, namely the known hadron spectrum has very few complete
$SU(3)$ multiplets, and recent LQCD calculations show the existence of
yet unobserved states, albeit at larger quark masses for
baryons~\cite{Edwardsj}. A first estimation of the missing states is based
on SU(3) and the PDG listed baryons is given by the number of different
strangeness isospin multiplets, namely: $\#\Sigma=\# \Xi=\# N+\#\Delta$
(PDG- 26; 12; 49), $\#\Omega=\#\Delta$ (4; 22), and $\#\Lambda=\#N+ \#
\text{ singlets}$ (18; 29). Thus on this count alone we are missing the
following isospin multiplets: 23 $\Sigma$, 11 $\Lambda$, 37 $\Xi$ and 18
$\Omega$. One expects even more missing states according to the quark
model, LQCD, and/or  the $SU(6)\times O(3)$ organization of multiplets.
The question is therefore how sensitive is the HRG to those missing
states, which consist in particular of a large number of hyperons.

The HRG is determined by the pressure, where the contribution to the
partial pressure by a given iso-multiplet $i$ is:
\beq
	p_i=T\frac{\partial}{\partial V}\log Z_i=T^2\, m_i^2 \,d_i \,
	\frac{1}{2\pi^2}\sum_{k=1}^{\infty}\frac{(-1)^{(1+k)B_i}}{k^2} \;
	K_2(k\frac{m_i}T)\;e^{k\mu_i/T} ,
\eeq
where $B_i$ the baryon number, $d_i=(2I_i+1)(2 J_i+1)$, and $\mu_i$
is a chemical potential, and $K_2$ is the modified Bessel function.
For our purposes where $T<0.16 \text{ GeV}$, keeping only the first
term in the sum is sufficient for all hadrons except the $\pi$, $K$
and $\eta$ mesons. Here we use the meson resonances listed in the PDG,
and for baryons we choose to use $SU(6)\times O(3)$ multiplets with
the mass formulas provided in Ref.~\cite{GMj}, where we will include
the $\bf{56}$ and $\bf{70}$ multiplets with $\ell=0,\cdots 4$.

\item \textbf{HRG in Chemical Equilibrium: LQCD}

We use here the results for QCD thermodynamics obtained in LQCD, and
the results are those of Ref.~\cite{Bazavov1j,Bazavov2j,Bazavov3j}. Above
the cross over transition there is a slow evolution towards the ideal
quark-gluon gas. Below the transition the HRG gives a remarkably good
description of the thermodynamic observables, 
The
figure also shows the effect of excluding  baryon resonances; those
effects for the pressure $p$ and the entropy $s$ are modest  for
$T<0.15$~GeV, and the effects of the hyperon resonances become almost
insignificant.  
Clearly it is not possible to disentangle the baryon resonance
effects through the global thermodynamic observables of the hadron gas
vis-\`a-vis the LQCD results. In chemical equilibrium, resonances rapidly
disappear with the falling temperature and so do their effects on total
thermodynamic observables.
It is therefore necessary to have  more sensitive observables  in order
to find the composition of the hadron gas:  for hyperons one needs  to
filter strangeness. This is achieved via the study of
correlations. In particular the susceptibilities (see for instance
Refs.~\cite{Rattij,corrBellwiedj,ERAj})  provide a useful tool. They are
defined by:
\beq
	\chi_2^{QQ'}\equiv \frac {1}{T^2}\frac{\partial^2 p}{\partial
	\mu_Q\partial \mu_{Q'}},
\eeq
where $Q$ and $Q'$ are conserved charges. If we consider only baryons,
we have that $\chi_2^{BB}\sim (n_B+n_{\overline B})/T^3$ and $\chi_2^{BS}
\sim (n_Y+n_{\overline Y})/T^3$: the HRG gives a very simple relation of
the susceptibilities to the particle number densities, which can be 
tested with the LQCD~\cite{Bazavov2}
The agreement 
is reasonably good, in particular for $\chi_2^{BS}$, which provides
perhaps the best indication of the role of excited hyperons for
$T>0.13$~GeV. For more extensive discussions of fluctuations and LQCD
results see~\cite{Rattij,Bellwied1j,ERAj}.

Although one expects resonances to have contributions whose magnitude
is similar to the ones estimated with the HRG, it is also true that
deviations from the approximation of the HRG may be of similar
significance, and thus the LQCD results do not seem to permit for a
definite estimate of what we are missing in terms of baryon resonance
states.

\item \textbf{HRG off Chemical Equilibrium: RHICS}

The hot hadronic system produced in high energy HICs is for most of
its brief expansion off chemical equilibrium. In the HRG description,
this requires the inclusion of chemical potentials  to account for
the overabundance of the different hadrons. The presentation here is
basic, ignoring possible effects of hydrodynamics, and corresponds to
describing the thermodynamics of the HRG in the local co-moving frame;
it should be reasonably good for discussing particle yields. In the
absence of net B, S and Q, we associate to each isospin multiplet a
chemical potential, equal to that of the corresponding antiparticles.
Processes which remain in equilibrium  give relations between chemical
potentials, e.g., if $A+B\leftrightarrow C+D$ is in equilibrium, then
$\mu_A+\mu_B=\mu_C+\mu_D$, or $A+B\leftrightarrow C^\ast$ implies 
$\mu_{C^\ast}=\mu_A+\mu_B$. The existence of different reaction channels   
requires information about partial rates. Such information is extremely 
poorly known for most resonances, in particular baryons, and therefore 
one needs to resort to models. For the purpose of our discussion we 
adopt a very simple model, which assumes: 
i) resonances are in chemical equilibrium with respect to their decay 
	products, 
ii) non-strange meson resonances decay only into pions, 
iii) strange meson resonances decay into one Kaon and pions, 
iv) all baryon resonances have 2-body decays into the ground state 
	octet and decuplet, 
v) all non-strange baryon resonances decay only via pion emission, 
vi) $\Sigma$, $\Lambda$ and $\Xi$ resonances decay with different rates 
emitting pions and K or $\overline K$. Resonance chemical potentials are 
then given by:
\bea
	\mu_i^{M^\ast}=\sum_{j=\pi,K} \nu_{ij}^{M^\ast} \mu_j~~,&~~~&
	\mu_i^{B^\ast}=\sum_{j=N,\Sigma,\Lambda,\Xi,\Omega}
	\nu_{ij}^{B^\ast}(\mu_j
	+\delta_{S_i\,S_j}\,\mu_\pi+\delta_{S_i\,(S_j\pm1)}\,\mu_K),
\eea
where the decay rates are encoded in:
\bea
	\nu_{ij}^{M^\ast} &=&\delta_{S_i0}\;\delta_{j\pi} \;\overline{\eta}_i
	+\delta_{S_i\;\pm 1}\,(\delta_{jK}+(\overline{\eta}_i-1)\delta_{j\pi})
	\nonumber\\
	\nu_{ij}^{B^\ast}&=&\delta_{S_i0}\;\delta_{S_j 0} +\delta_{|S_i| 3}\;
	\delta_{|S_j| 2}+\sum_{S=1,2} \delta_{S\; |S_i|} \,(r_i \,\delta_{S_i\;
	S_j}+\frac{1-r_i}{2}\,\delta_{S_i\,(S_j\pm 1)})~~,~~~
	\sum_j  \nu_{ij}^{B^\ast}=1,
\eea
where $\overline{\eta}_i$ is the average particle multiplicity in
the decay of the meson resonance $i$, and $r_i$ is the branching fraction
of pion emission in the decay of the baryon resonance $i$. With this one
can define effective particle number densities:
\bea
	\overline{n}^M_i= n^M_i+\sum_j \nu^{M^\ast}_{ji} n_j^{M^\ast}~~,&~~~&
	\overline{n}^B_i= n^B_i+\sum_j \nu^{B^\ast}_{ji} n_j^{B^\ast},
\eea
where $i$ indicates a stable meson or baryon, and $n$ are the number
densities obtained with the corresponding chemical potentials. For
simplicity we have neglected the baryon resonance decay contribution to
$\overline{n}^M_i$. The particle number densities  $\overline n_i=\partial
p/\partial\mu_i$ are the ones observed after kinetic freeze out.

Here the discussion is focused on the possible effects of resonances in
the observed particle yields in RHICs. One can consider ratios of yields,
and also fluctuations. The ratios remain constant after chemical freeze
out. One can consider more detailed chemical freeze out for different
hadrons~\cite{Bellwied:2016kpjj}, but for brevity we take a simple one,
namely freeze out at about $T=0.15$~GeV for all stable hadrons. Due to our
lack of knowledge and for simplicity we take $r_i=r$ for all $i$, and
check the dependencies on $r$. Using the yield ratios with respect to
pions from ALICE~\cite{ALICEj}, 
we choose $\mu_\pi=0$ at
$T=0.15$~GeV (this choice is  arbitrary here, and in particular the
value of $\mu_\pi$ at kinetic freeze out  is sensitive to it);
determinations of the initial $\mu_\pi$ can be improved via knowledge of
it at kinetic freeze out. Fitting to the ALICE yield ratios one fixes the
chemical potentials $\mu_i$ at the that initial $T$. To see how the HRG
evolves one uses the approximation~\cite{BGGLj} that the ratios of the
densities of conserved particle numbers to total entropy remain
approximately constant, namely $\overline{n}_i/s\sim\text{const}$.
One notices that,
as expected, the baryon resonances play a very marginal a role in the
evolution of the meson component of the HRG (mostly because of the
assumption in Eqn.~(5)); for nucleons the effects are also marginal
except at higher $T$. The main sensitivity to resonances is on the
hyperons,  and it depends  very much on the value of $r$. The effects
of baryon resonances on the effective chemical potentials defined by the
relation $n_i(T,\overline{\mu}_i)=\overline{n}_i(T,\{\mu_j\})$ are 
also relatively small. Clearly the effects of the baryon resonances 
are below other effects, such as the initial conditions right after 
hadronization and/or the uncertainties in the resonances' partial decay 
fractions. Thus, it is necessary to use more sensitive observables to 
acquire more sensitivity. For that purpose, one can consider 
susceptibilities such as the ones mentioned earlier, now adapted to a 
HRG off chemical equilibrium. They are simply defined by:
\beq
	\chi_2^{ij}=\frac{1}{T^2}\,\frac{\partial^2 p}{\partial \mu_i 
	\partial \mu_j},
\eeq
where the experimentally accessible quantities are ratios of those
susceptibilities, namely:
\beq
  R^{ij}_{kl}=\frac{ \chi_2^{ij}}{ \chi_2^{kl}}.
\eeq
Using the HRG off chemical equilibrium one easily calculates the ratios.
The ratios
are sensitive to the value of $r$: while for the case $r=1$ one cannot
distinguish effects of resonances, those effects are  significant for
$r=0.5$. In particular one needs the resonances to have a non-vanishing
$R^{\Omega\Xi}_{NN}$. Thus, the analysis of particle number fluctuations
and correlations are the most sensitive tool to extract information on
resonance effects from RHICs data.  Those effects come however modulated
by resonance partial decay rates which are little known. For further 
discussion  involving higher order correlations/fluctuations see  
Ref.~\cite{BellwiedProcj}.

\vspace{1cm}
\item \textbf{Comments}

In principle, it is possible to obtain indications of hadron resonance
effects, in particular baryons, in the hadronic phase of hot QCD. Studies
with LQCD and RHICs are the sources of relevant information. Using the
HRG one can then quantify those effects within the model, and draw
conclusions as discussed in this note. For the case of LQCD, the main
sensitivity to resonances resides in fluctuations/correlations  such as
susceptibilities, which strongly indicate the importance of hyperon
resonances for $T>0.13$~GeV. In the case of RHICs there is less
certainty on the conclusions due to lack of knowledge of the initial
thermodynamic state of the hadronic  fireball and  of  the partial
decay rates of resonances which affect the  expansion off chemical
equilibrium. There is however interesting information encoded in
ratios of correlations, which may definitely require the effect of
resonances to be explained.

\item \textbf{Acknowledgments}

This work was supported in part by DOE Contract No. DE--AC05--06OR23177 
under which JSA operates the Thomas Jefferson National Accelerator 
Facility and by the National Science Foundation through grants 
PHY--1307413 and PHY--1613951.

\end{enumerate}


\newpage
\subsection{The UVa approved and proposed experiments}
\addtocontents{toc}{\hspace{2cm}{\sl D.~Keller}\par}
\setcounter{figure}{0}
\setcounter{table}{0}
\setcounter{equation}{0}
\setcounter{footnote}{0}
\halign{#\hfil&\quad#\hfil\cr
\large{Dustin Keller \footnote{Email:  dustin@jlab.org}}\cr
\textit{University of Virginia}\cr
\textit{Physics Department}\cr
\textit{U.S.A.}\cr}

\begin{abstract}
Compton Scattering, though one of the most fundamental processes, is still
not well understood at the intermediate (Jlab) energies.  There maybe a
critical link in this process to the General Parton Distributions
(GPDs) but there is still considerable disagreement within the leading
theoretical frameworks.  A novel experimental configuration using a high
intensity photon source (HIPS) and rotating target at Jlab is discussed.
The measure the initial state helicity correlations of real Compton scattering
at wide angles is flagship experiment for many HIPS experiments to come.
Some aspects are given for a likely design of the rotating target raster.
\end{abstract}

\begin{enumerate}

\item \textbf{Introduction}
Some significant progress has been made over the last decade in our understanding
of exclusive reactions in the hard scattering regime.
This progress had been made possible (in part) by data from Jefferson Lab
on elastic electron scattering and Compton scattering from the proton
and by a significant and increasingly sophisticated theoretical effort
to exploit the richness of exclusive reactions at moderate momentum transfers.

Results of experiments at Jefferson Lab on the proton contradict
the predictions of pQCD: the recoil polarization measurements of \gep\, E93-027, E04-108 and E99-007,
and the Real Compton Scattering (RCS) experiment E99-114.
The \gep~measurements~\cite{jo00,ga02,ap10} found that the ratio of $F_2$ and $F_1$,
scaled by $Q^2$ demands a revision of one of the precepts of pQCD, namely
hadron helicity conservation.
Results from the RCS measurements~\cite{Hamilton:2004fq,dham} are that the longitudinal
polarization transfer \KLL~is large and positive, contrary
to the pQCD predictions for \KLL.  These experiments provide a compelling argument that pQCD should not be
applied to exclusive processes at energy scales of 5-10 GeV.

An alternate theoretical framework exists for the interpretation of exclusive scattering
at intermediate energies~\cite{ra98,di99,hu02,ca02}.
This alternative approach asserts the dominance of the handbag
diagram in which the reaction amplitude factorizes into
a subprocess involving a hard interaction with a {\it single quark}.
The coupling of the struck quark to the spectator system is described by
the Generalized Parton Distributions (GPD's)~\cite{ji97,ra96}.
Since the GPD's are independent of the
particular hard scattering reaction, the formalism leads to a
unified description of hard exclusive reactions.  Moreover, the
relationship between GPD's and the normal parton distribution
functions provides a natural framework for relating inclusive and
exclusive reactions.

SCET~\cite{ki13} has been used to develop a description of the soft-spectator scattering contribution in a complete factorization
for the leading power contribution in wide angle Compton scattering.
SCET attracted considerable attention recently for its QCD factorization approach of the two-photon exchange (TPE) contributions
to elastic electron-proton scattering.  From the TPE development a single universal SCET form factor was found to define the dominant soft-spectator
amplitudes which auspiciously is the same form factor needed to describe complete SCET factorization in WACS.

The GPD approach~\cite{di13} has a relatively evolved interpretation of the framework.  In the GPD model there are already predictions for much
of the Compton polarized observables and ways to relate these observables to one another in particular kinematic regions providing additional
model constraints.  There are still necessary assumptions of restricted parton virtualities and power corrections but the chance to expand the experimental
tools and methods to access GPDs is inviting.  In this framework RCS have a complementary nature to DVCS in so far as in DVCS the GPDs are probed at small $t$ while for RCS (and nucleon form factors) the GPDs are probed at large $t$.

An additional handbag approach~\cite{mi04} attempts to incorporate the influence of quark transverse and orbital angular momentum and the corresponding violation of proton helicity conservation.  Predictions for the cross sections from this model are found to be in reasonably good agreement with early measurements. The helicity correlation between the incident photon and outgoing proton, $K_{LL}$, is found to be in the same approximate region over the center of mass angle as the other handbag predictions.  However, $A_{LL}$ deviates significantly from $K_{LL}$ and infact even changes sign for large angles.

The high intensity photon source that will be used for this experiment is still under intense study for optimization.  Some preliminary planes can be found in proposals \cite{pro1, pro2} and these proceedings.

\item \textbf{Goals of the Experiment}

\label{eval}
Taking data of the initial state helicity correlations at large Mandelstam variables where the theoretical framework in
under control and came be interpreted is essential.  This data can be used to study how the RCS reaction proceeds with
the interaction of photons with a current or constituent quark to investigate the hard and soft contributions.  The data
so far taken does not give confidence that our understanding of the polarized observables is clear.  This means it is necessary to
build on data of the polarized observable to assist in the interpretation of the phenomenology required to move the theoretical
framework forward.  If the GPD model proves to be useful the data should be taken to help to determine the Compton form factor ratios
by measuring the polarized target asymmetries over a range of kinematics.  This experiment can provide the maximum amount of information and constraints on the theory by holding some kinematics variables fixed studying the variation in others.  The focus of the experiment
can be summarized as follow,
\begin{itemize}
\item Take $A_{LL}$ Data to compare with the already measured $K_{LL}$: {\small Helps to build understanding of contributions from orbital angular momentum, checks the GPD relationship extracted from form factors through the sum
rules, help to provide needed information on the hard subprocess, current/constituency, proton helicity flip contribution.}

\item Take $A_{LS}$ Data to compare with $K_{LS}$ : {\small Study form factor ratio through a new
observable and checking previous measurements.  Most powerful discrimination between GPD and SCET approach.  Tests
fundamental predictions in the competing interpretations.}

\item Data at Large angle, Large Mandelstam Variables : {\small Large angles provides the best discrimination
between the handbag approaches and pQCD.  All kinematics should have $|t|$ and $|u|$ above 2.5 GeV$^2$ where the handbag can be applied.}

\item Fix $s$ and study change in $\theta$, fix $\theta$ and study change is $s$ : {\small For both GPD
model s-dependence can be measured but for SCET s-dependence is small this gives a way to discriminate between the two as well as providing the needed information to parameterize the GPD approach.}

\item Expand number of measured observables : {\small To help theory parameterize and interpret phenomenology.  The more polarized observables that can be measured, the more information is provided that can be use to constrain theory as it is clear predictions are so far very limited in predicting data.}
\end{itemize}

\item \textbf{Uniform Illumination of the target cups}

\begin{figure}[!htbp]
   \centering
   \includegraphics[width=1\textwidth]{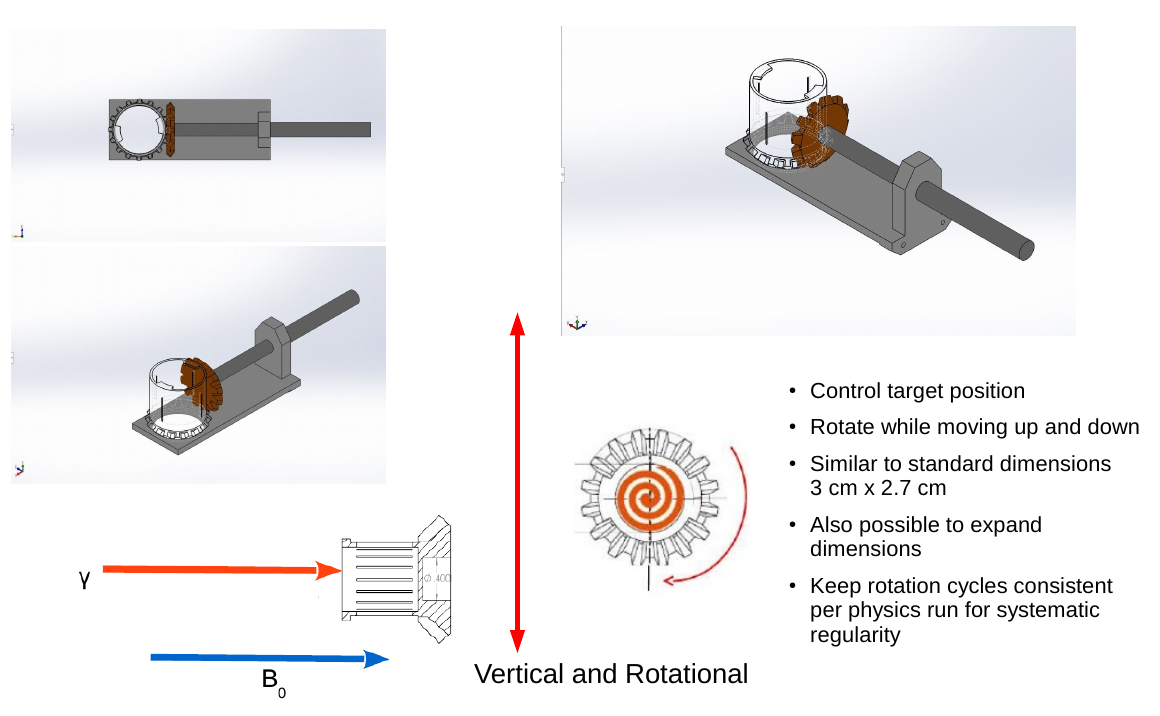}
   \caption{A simple geared cup example specialized so that the target cup does not interact with the beam.  Vertical motion combined with rotation of cup will allow uniform coverage of target cell. The red dot represents the fixed position of the photon beam. The colored bead in the cup can be seen moving as the cup rotates counterclockwise and the target ladder is moved up.}
   \label{fig:raster1}
\end{figure}
Though a dilution refrigerator is quite common with pure photon beams, the low cooling power of this type of system is not adequate for HIPS.  An evaporation refrigerator with polarized NH$_3$ is the necessary polarized proton target for a pure photon beam of the order of $10^{11}$-$ 10^{12}$ $\gamma/s$.  For electron beam experiments typically 100 nA is the maximum current on the target which leads to approximately 0.36 W heat load.  Combined with the heat deposit from the microwave, used to dynamically polarize the target, the cooling power of the UVA/JLab evaporation fridge is near saturation.  For any of the JLab HIPS configurations which produces near the mentioned photon flux using a $\sim$10\% copper radiator in combination with a photon beam collimator an electron beam current of  $\sim$$1 \mu$A is required.  Simulations indicate that beam current above 3 $\mu$A  starts to become difficult to manage in regards to maintaining minimal radiation near the magnet and target area using a local dump.  Nevertheless if cooling power was the only limiting issue one could expect to run at more than $10 \mu$A ($\ge 10^{13}$ $\gamma/s$) before putting any noticeable heat load on the cooling power of the refrigerator.  At the scale of $10^{12}$ $\gamma/s$ effects from the heat load of the beam seen in electron beam experiments, such as jumps in polarization due to beam trips or current changes, would be near negligible.  This can improve the systematics related to polarization calculations in the data analysis.

Cooling power is not the only concern.  Solid polarized targets suffer from radiation damage and local hots spots can cause depolarization of the target.  A focused HIPS on the scale discussed (RMS $< 1$ mm) would lead to depolarization in the region of the photon beam spot with the greatest intensity due to material interfacial thermal heating and radiation damage.  
  To achieve acceptable resolution a beam spot with a RMS of less than a millimeter is desired, resulting in a very intense photon spot on the target.  Photons at the several GeV range can easily brakeup the NH$_3$ to make additional paramagnetic centers in the target that can adversely effect the maximum achievable polarization.  The production of these free radicals happens in electron beam experiments as well and is the leading cause for target maintenance and the overhead time required to anneal or replace the target material.  The full composition of the paramagnetic complex in an exhausted material is not well known but the over all degradation behavior as a function of dose is understood from empirical studies for $\sim$100 nA in Hall A and C as well as $\sim$10 nA in Hall B.  It is also know that the ionizing electron beam can cause more than an orders of magnitude more radiation damage at the same beam intensity as photons.  But there are also other aspects of the photon beam target interaction that are important to consider.  The photon beam causes bremsstrahlung radiation in the target and electron positron pairs.  This secondary scattering of ionizing radiation inside the target can also cause radiation damage.  Using simulations for a photon flux of $\ge 10^{11}$ $\gamma/s$ with RMS$\sim$1 mm leads to the resulting equivalent in electron beam current from the secondary scattering of ionizing radiation to be $\sim$20 nA at the exit of the target in an area of 4.5 mm$^2$.  If this dose can be distributed over the full surface of the target such that the same dose is distributed over an area of 570 mm$^2$ then the expected depolarization due to radiation damage begins to approach the solid polarized NH$_3$ behavior seen in Hall B.  This considerably reduces the number of anneals and target changes over the course of the experiment as compared to the standard UVA/JLab operation in Hall A and C.  The decay of the polarization is calculated to be about 5 times slower over the course of the experiment leading to improvements in the figure of merit for the same requested experimental run time.
\begin{figure}[!htbp]
   \centering
   \includegraphics[width=.35\textwidth]{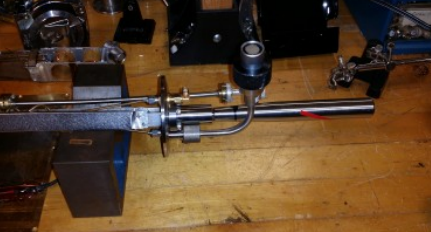}
   \caption{The rotary feed-through at the top of the target insert used to drive the rotation of the geared cup.}
   \label{fig:raster2}
\end{figure}
In order to increase the area of the target that the photon beam will interact with a rotating target was developed to raster photons over the target cup face, see Fig. \ref{fig:raster1}.  The Kel-F target cup is machined to include a gear that can be driven from a rotating shaft along the target insert.  Fig. \ref{fig:raster1} shows a design of a similar dimension to polarized targets used in the past that fill up the entire homogeneous field region of the polarizing 5 T magnet.  In the design shown there is no additional material from the cup in the beam-line.  In the design shown there is no additional material from the cup in the beam-line.  The front and back of the target cell are made of a thin Aluminum foil.  The NMR couples inductively to the target material by a coil wound around outside of the cup.  The rotating shaft passes through the top of the target insert using a vacuum rotary feed-through, shown in Fig. \ref{fig:raster2} which is then driven by a electric motor. 

The target rotation in combination with the standard target actuator results in an effective slow raster which spirals over the full area of the standard 2.7 cm diameter target.  The beam collimatimation provides the spot size on the target and couples directly to the resolution characteristics for reconstruction at the cost of holding the beam location in space fixed.  We can still obtain uniform exposure of the target cell by a combined rotation of the target cup synchronized with an up/down movement of the target ladder, see Fig.~\ref{fig:raster1}. Rotation of the target cup has already proven viable in may UVA tests.  An example of one of the many rotating cups used at UVA is shown in Fi.g \ref{fig:raster3}.  Depolarization and inhomogeneous radiation damage can easily be achieve by continuously moving the target at a rate determined by the radius of the circle made through rotation on the target surface, spending no more than a few hundred milliseconds on each target location.  So even near the center only $\sim$0.01 Hz is required.  At UVA rotation rates of several Hz have been used.  By completing a fixed number of rotations for each experimental run, fluctuations from the variations in target bead packing can be averaged out.

\begin{figure}[!htbp]
   \centering
   \includegraphics[width=.3\textwidth]{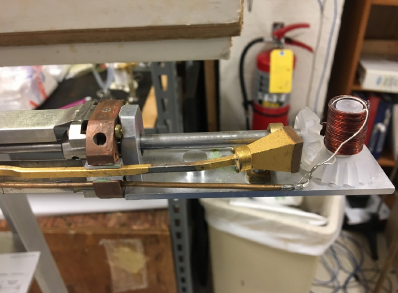}
   \caption{One of the many rotating cups used at UVA, this is a special design for small targets.}
   \label{fig:raster3}
\end{figure}

\item \textbf{Conclusion}
The polarized observables of Compton Scattering are much needed to
provide additional constraints on theory to obtain additional information 
on the reaction mechanism at the intermediate energy.  A high intensity
photon source with a pure photon allows for an Compton interaction rate that
allows higher energy making essential tests possible in the kinematic region that the handbag
approaches can be tested. A novel experimental configuration using a HIPS and rotating target at Jlab makes
polarized WACS highly feasible.
The measure of the initial state helicity correlations of real Compton scattering
at wide angles will be a flagship experiment for many polarized target HIPS experiments to come.

\item \textbf{Acknowledgments}
This work was supported by DOE contract DE-FG02-96ER40950.

\end{enumerate}


\newpage
\subsection{Large Acceptance Wide-angle Compton Scattering Experiment}
\addtocontents{toc}{\hspace{2cm}{\sl B.~Wojtsekhowski}\par}
\setcounter{figure}{0}
\setcounter{table}{0}
\setcounter{equation}{0}
\setcounter{footnote}{0}
\halign{#\hfil&\quad#\hfil\cr
\large{Bogdan Wojtsekhowski\footnote{Email:  bogdanw@jlab.org}}\cr
\textit{Thomas Jefferson National Accelerator Facility}\cr
\textit{Physics Division}\cr
\textit{Newport News, VA 23606, U.S.A.}\cr}

\begin{abstract}
The conceptual ideas for a high productivity measurement of a double spin asymmetry of
Wide-angle Compton Scattering from a proton are presented.
\end{abstract}

\begin{enumerate}
\item \textbf{Introduction}

There is large interest in accurate measurement of the double spin asymmetries
of Wide-angle Compton Scattering from a proton at the invariants $s, -t$, and $-u$, 
all to be much larger than $\Lambda_{_{QCD}}^2 \sim$ 1 GeV$^2$.
The measurements should be focused on the most urgent issues and provide input for 
advancing our knowledge of hadron physics.
The physics of WACS is not fully understood, which is especially striking when we compare it
with the level of understanding of other reactions considered in the same GPD framework.
The dominant reaction mechanism was identified about 15 years ago after advances
in QCD phenomenology with GPD formalism~\cite{ra98,di99,hu02} and the first precision measurements
of the $K_{_{LL}}$~\cite{ha05} and the cross section~\cite{da07}.
However, a recently completed~\cite{fa15} second measurement of $K_{_{LL}}$  disagrees with published predictions.
A test of the theory predictions~\cite{hu02,ki13} requires new measurements at higher incident photon energy. 
It is especially important to measure the $K_{_{LL}}$ (or $A_{_{LL}}$) observable in 
a wide range of photon energy at $90^\circ$ in a center-of-mass (cm) system where the theory predictions are 
fully applicable and experimental results will provide important constraints on GPD models.

\item \textbf{Design a productive double polarization experiment}

The design of an advanced experiment requires an optimized combination of many variables.
Among them,  the most important are the detector acceptance, luminosity, and effective polarization.
We will skip a comparison of the $K_{_{LL}}$ vs. $A_{_{LL}}$ approaches to double polarized WACS
and proceed with our formulation of the optimized $A_{_{LL}}$ experiment.

There are two detectors in the WACS experiment: a calorimeter for measurement of
the scattered photon energy and coordinates at the detector, and a magnetic spectrometer for
measurement of the recoil proton momentum and direction of motion.
The combined acceptance requires optimized detector locations. 
At a $90^\circ$ in cm system for a two-body final state coincidence experiment, the solid angles and
the vertical-to-horizontal  aspect ratio of the angular acceptance should be similar in the two arms.
This is largely achievable for the $90^\circ$ cm case because of a good coordinate resolution of the
calorimeter and flexible distance between the target and the calorimeter.
The usefulness of the very large detector acceptance is limited by a change in the
counting rate within the acceptance and the maximum luminosity at which a specific detector could operate.
However, the SBS horizontal acceptance of $\pm 5^\circ$ (in the lab) is sufficiently small.

For the measurement of these polarization observables we can use 
a wide angular acceptance because these observables for WACS do not change fast. 
The range of $\theta_{cm}$ of $\pm 10^\circ$ corresponds to a variation of $t$ by $\pm 12$\% 
and a similar change of $A_{_{LL}}$, and the counting rate changes from the low side to the 
high side of the $t$ range by a factor of 2.5.
The momentum acceptance of the proton arm together with the detector acceptances 
constrains the energy range of an incident photon.
With the SBS acceptance, the $s$-range could be as large as 50\% if WACS event selection 
is sufficiently clean.

The luminosity of the experiment is defined by the intensity of the photon beam and 
the parameters of the polarized target.
Due to the limitation on the heat load for a low temperature NH$_3$ target and 
the rate of radiation damage of the material in the target, 
the beam intensity of $5-8\times10^{12}$ equivalent photons per second is close to an optimum,
but even a three to four times higher intensity could be useful if the time overhead for
the target annealing could be reduced.
The production of such an intense photon beam requires an electron beam of significant intensity.
For a 10\% radiation length converter, an electron beam of 1.2~$\mu$A is needed, which leads to
the corresponding electron beam power of 10~kW.

The quality of the results depends very significantly on WACS event selection and low pion dilution, 
which could be achieved thanks to an exact angular correlation in the two-body final state of the WACS process
but requires a good angular resolution in both arms of the detector apparatus.

There is interest in a test of the $A_{_{LL}}$ and $K_{_{LL}}$ equality obtained for WACS in the GPD approach.
The test is motivated in part by a potential contribution of a constituent quark effect~\cite{mi04}.
We noticed that a similar equality was also obtained for a single pion photo-production case~\cite{hu04}.
The high precision of $K^\pi_{_{LL}}$ data points~\cite{ha05,fa15} allows us to do a productive check
of the theory prediction because the corresponding $A^\pi_{_{LL}}$ data could be obtained quickly
and accurately using a mixed electron-photon beam approach which provides the means for 
reliable control of systematic uncertainty and good statistical precision.

\item \textbf{The Hermetic Compact Photon Source}

A full description and the analysis of the HCPS are presented in the contribution~\cite{HCPS} of these proceedings.
Here, we only explain why HCPS is dramatically better than a traditional scheme with the greatly
separated locations of the converter and the beam dump.
The root of the answer is in the large tail of the electron energy spectra after it passes through a converter
of 10\% radiation length.
When the beam dump is constructed at large distance from the converter area, 
the opening of the dump needs to be very wide because of the huge beam size, 
which leads to a large size and high cost of the dump in addition to the radiation shielding of 
the converter area, where the power deposited is only a few times lower.
The traditional scheme works well for a very thin converter of $10^{-4}$ radiation length, which 
is typical for the tagged photon facilities, but for a high intensity source practical results 
were not impressive due to high radiation background in the experimental hall.
Importance of the tail is easy to confirm with a simple calculation of the power deposition.

Indeed, the power loss in the converter area could be easily calculated as the sum of the power in 
the electron energy spectra tail and the power loss in the collimator of the photon beam. 
The power in the tail is $P_b \times t_c \times \left [ \ln (E_b/E_{cut}) -1 \right ]$, where $P_b$ is
the power of the incident electron beam, $t_c$ is the converter thickness in units of radiation length, $E_b$ is the beam
energy, and $E_{cut}$ is the acceptance of the beam dump (the maximum energy loss for which 
an electron will be able to reach the beam dump).
For a typical ratio $E_{cut}/E_b \sim 0.03$ and a 10\% converter, the dump will recieve three times more 
power than the converter area even if the photon beam loss due to collimation is ignored.
At the same time, the additional radiation shielding required for the 90\% of the beam power vs. 25\% is modest
and could be achieved by an extra twenty cm of concrete or five cm of tungsten.

\item \textbf{Large acceptance for the polarized WACS measurement}

The angular acceptance of the Super Bigbite spectrometer is 70 msr with an aperture ratio 
of 2.6:1 (vertical to horizontal).
This aspect does not match perfectly with the NPS's almost square aspect ratio of 1.2:1. 
However, the excess of the SBS acceptance helps us to measure the shape of pion photo-production
event distribution, whose intensity and the asymmetry $A^\pi_{_{LL}}$ 
need to be taken into account precisely.
The overall acceptance of the proton arm for 7.5~GeV incident photon energy was found to be 39~msr,
which is 6-7~times larger than with any universal magnetic spectrometer.
In addition to a large angular acceptance, 
SBS has the advantage of a large momentum acceptance (unlimited above 2~GeV/c)
and also superior angular resolution (essential for WACS event selection). 

As in the previous experiments, an aerogel-based Cherenkov counter should be used
for rejection of the charged pion events in the proton arm.
The momentum acceptance allows collecting of the proton-photon events for initial photon energy 
well below the beam energy and facilitates the photon energy ``$s$-scan" with one setting 
of the apparatus and beam energy.

The projected accuracy for double spin asymmetry $\sigma_{_{A_{_{LL}}}}$ is~0.08 at $s \,=\, 15$~GeV$^2$
and would be achieved with 15 days of data production time.

\item \textbf{Acknowledgments}
The author would like to thank the organizing committee of HIPS2017  for the invitation to the meeting. 
I greatly appreciate discussions with the workshop participants, the members of NPS collaboration, 
and many years of fruitful collaborations with P.~Kroll, P.~Degtyarenko, D.~Hamilton, and G.~Niculescu.
This work was supported by contract number DE-AC05-06OR23177, under which 
the Jefferson Science Associates operates the Thomas Jefferson National Accelerator Facility.

\end{enumerate}


\newpage
\subsection{Wide-angle exclusive photo-production of $\pi^0$ mesons}
\addtocontents{toc}{\hspace{2cm}{\sl Simon \v{S}irca}\par}
\setcounter{figure}{0}
\setcounter{table}{0}
\setcounter{equation}{0}
\halign{#\hfil&\quad#\hfil\cr
\large{Simon \v{S}irca \footnote{Email:  simon.sirca@fmf.uni-lj.si}}\cr
\textit{Department of Physics}\cr
\textit{University of Ljubljana}\cr
\textit{Slovenia}\cr}

\begin{abstract}
This talk represented a status report on the TJNAF E12--14--005 Experiment [1]
approved to run in Hall C of Jefferson Lab.  This experimental effort
is motivated by the fact that hard exclusive reactions are ideal for
studying hadron dynamics of underlying parton-level processes and
thereby allow one to test scaling relations, investigate the short-range
structure of nucleons and, possibly, explore the onset of transition
from non-perturbative to perturbative QCD.
\end{abstract}

\begin{enumerate}
\item \textbf{Introduction}

Existing data on the differential cross-sections d$\sigma$/d$t$
for the $\gamma p\to p \pi^0$ process 
as functions of $s$ and $|t|$ have been shown at various
center-of-mass angles $\theta^* = 50^\circ$, $70^\circ$, $90^\circ$
and $110^\circ$, including the latest g12 (CLAS) results from 
$\gamma p\to p\pi^0\to p\gamma\gamma \to pe^+e^-\gamma$.
It appears that at high $s$ and large $\theta^*$ the cross-sections
are consistent with $s^{-7}$ scaling ($s^{-n}$, $n = 6.89 \pm 0.26$).
It has been shown in the past that charged photo-production data 
at highest $s$ also indicate scaling, a feature which one would now
like to observe in the neutral channel and at much larger $s$ than so far.
These preliminary also reveal that the handbag (GPD-based) description
--- which operates on the assumption that the hard part of the process
(photo-production of the meson off a quark) and the soft part of
the process (extraction of a single quark from a hadron and its
re-absorption) factorize --- is insufficient and typically 
under-predicts the data by several orders of magnitude.  
On the other hand, Regge-type approaches appear to describe
the data well, at least qualitatively and in terms of functional forms.

One way to search for signatures of the handbag mechanism is also
to form the ratios of photo-production cross-sections.  This has
been attempted in the charged channels where the 
${\mathrm{d}\sigma(\gamma n\rightarrow\pi^-p)}/
{\mathrm{d}\sigma(\gamma p\rightarrow\pi^+ n)}$ ratios
were shown to agree well with the data, while such agreement
is lacking in the neutral case.  One of the reasons for this could
be the cancellation (or non-cancellation, respectively) of the form-factors 
of the quarks involved in the soft part of the process.  The talk 
has also touched upon the various explanations of why asymptotic scaling 
might be broken and what could be the reason for oscillations 
(or lack thereof) in the photo-production cross-sections 
at low (high) energies.

The setup of the proposed experiment has been described, together with
the illustration of the kinematic coverage, noting that the data taking
will be concurrent with the Wide-Angle Compton Scattering (WACS) experiment
[2] utilizing the same equipment.  The coverage will be approximately
$9.5 \, \mathrm{GeV}^2 < s < 21 \, \mathrm{GeV}^2$ and 
$60^\circ < \theta^* < 110^\circ$.  Event identification has been
discussed, for which there are two options: with single-$\gamma$ 
detection or by detecting both photons.  In the first case one
assumes two-body kinematics and uses the difference between the measured 
predicted hit position in the calorimeter to identify the reaction.
This method allows for a clean enough separation of single-$\pi$ 
from two-$\pi$ events as well as of $\pi$ and $\eta$ production,
resulting in typical contaminations of a few percent.
In the second case one constructs the missing-mass spectrum based
on the two detected photons with a peak around the pion mass
with typical widths of a few MeV.  Both methods have been shown to work
well in previous real-photon experiments, in particular WACS.

\end{enumerate}


\newpage
\subsection{Radiological Issues at JLab - Lessons Learned from the PREX Program}
\addtocontents{toc}{\hspace{2cm}{\sl R.~Beminiwattha}\par}
\setcounter{figure}{0}
\setcounter{table}{0}
\setcounter{equation}{0}
\setcounter{footnote}{0}
\halign{#\hfil&\quad#\hfil\cr
\large{Rakitha Beminiwattha \footnote{Email:  rakithab@latech.edu}}\cr
\textit{Louisiana Tech University}\cr
\textit{Physics Department}\cr
\textit{U.S.A.}\cr}

\begin{abstract}
In this talk the impact of high-intensity photon sources on radiological issues for experiment/equipment in the experimental hall is discussed, specifically for the PREX/CREX and MOLLER experiments at JLab.
\end{abstract}

\begin{enumerate}
\item \textbf{PREX-I}
PREX-I experiment used a beam intercepting collimator to block low angle scattered electrons and Bremsstrahlung photons. The collimator aperture was set at $\rm 1.27^o$ but the main limiting aperture, the scattering angle above which beam interact with the beam pipe was smaller than the collimator aperture and it was located at the downstream area of the septum beam pipe (the gate-valve ) at $\rm 0.84^o$. The electromagnetic power from the low angle scattered electrons and photons from the lead target interact at the collimator as well as at the limiting aperture. They acted as secondary neutron sources. A combination of fringe field leak from the septum magnet into the beam-pipe and the limiting aperture at septum area resulted in significant spray in to the hall and beam-pipe downstream of the septum area. This produced many secondary radiation sources along the beam-pipe in the hall A during PREX-I experiment. Due these issues, PREX-I produced more neutron radiation compared to any standard hall A experiment. 

\item \textbf{PREX-II}
The main strategy for PREX-II experiment is to use a single collimator to stop everything that misses the dump and act as the limiting aperture for the experiment. The collimator aperture for PREX-II is set to $\rm 0.78^o$. The new collimator will intercept about 2 kW of power at $\rm 70\ \mu A$ from the beam and it is expected to produce more neutrons. The PREX-I collimator only intercepted 500 W. The collimator is designed to self-shield high energy neutrons and slow down to more softer neutron spectrum (less than 10 MeV) using an outer tungsten jacket. The inner cylinder of the collimator which directly interact with the beam is made out of copper-tungsten alloy to provide good thermal conductance for better cooling. This inner core will be water cooled. Then a neutron shield is implemented around the collimator to reduce the soft neutron energy spectrum. The soft neutron energy spectrum is shielded using high
density polyethylene (HDPE) that provide full shielding around the collimator. Due to high intensity beam interactions the collimator will be heavily activated by the time PREX-II is completed. Therefore, the collimator will be retracted to a 5 cm thick lead shielded box before de-installation. 

\item \textbf{Impact of neutrons}
The main problem caused by neutron spectrum produced by PREX-I experiment was due to cumulative effects of displacement damages to Silicon in electronic devices. The atomic degradation overtime will ultimately proved destructive for electronics in the Hall A. Many power supplies and other electronics were damaged during the PREX-I experiments. One unique observation was that many optocoupler units were damaged more often than other electronics units. This was due to lower threshold for failure of optocoupler due to neutron cumulative effects. Therefore as a preventive measure all the optocoupler units were removed from the Hall A. Also PREX-II experiment will be designed to have about order of magnitude less neutrons produced compared to PREX-I experiment. 

\item \textbf{Conclusion}
In concluding, a combination of large aperture collimator and septum fringe created many neutrons sources in the hall during PREX-I. Isolating the main neutron source and then shielding adequately this main source is the optimum solution to minimizing the neutron radiation. This include use of self-shielding collimators and them use concrete and HDPE to shield neutrons. The studies have shown that collimation and shielding strategy reduces the expected radiation load in PREX-II to the level of previous successful experiments such as HAPPEX-2 or PVDIS in the most sensitive region of the hall. The simulation benchmark between Geant3, Geant4, FLUKA, and MCNPX have shown that agreement of neutron production is withing a factor of 2. Comparison of simulation results with RADCON data during PREX-I experiment have shown that there is a factor of 2 safety margin between simulation and measurements.


\end{enumerate}

%

\newpage
\subsection{DVCS and TCS in New Helicity Amplitudes Formalism}
\addtocontents{toc}{\hspace{2cm}{\sl S.~Liuti}\par}
\setcounter{figure}{0}
\setcounter{table}{0}
\setcounter{equation}{0}
\setcounter{footnote}{0}
\halign{#\hfil&\quad#\hfil\cr
\large{Simonetta Liuti \footnote{Email:  sl4y@virginia.edu}}\cr
\textit{University of Virginia}\cr
\textit{Physics Department}\cr
\textit{U.S.A.}\cr}

\begin{abstract}
I will summarize results using our new helicity amps. based formalism. The formalism allows us to connect in a more straightforward way the matrix elements for DVCS and TCS up to twist three. I will deal in particular with the issue of time reversal, and its consequences for GPDs universality.
\end{abstract}

\begin{enumerate}
\item \textbf{Introduction}
The cross section for exclusive deeply virtual photon electroproduction
contains a Bethe Heitler (BH), a pure DVCS and a BH DVCS interference term. Similarly, the cross section for exclusive deeply virtual lepton pair production can be written in terms of BH,  pure TCS and BH TCS interference contributions. 
For DVCS/TCS one can choose a kinematic setting where the leptons lie in a plane with the virtual photon along the z axis and  the final proton and final/initial photon form a separate plane -- the hadron plane -- at an angle $\phi$ (see {\it e.g.} Figure \ref{fig:kinematics} for DVCS). 
The DVCS/TCS cross section factorizes into a lepton part which is $\phi$ independent, and a hadron part which is $\phi$ dependent. For BH, the lepton part contains a photon which is emitted/absorbed by one of the two electrons, therefore it is also $\phi$-dependent. 

\noindent {\it  It  is this rather simple aspect of the kinematical setting for BH that gives origin to the  complicated angular dependence of the BH DVCS interference term. }

Several papers have been written to date which describe the cross section for DVCS and related processes (see \cite{Belitsky:2001ns,Belitsky:2005qn,Belitsky:2010jw} and \cite{Kumericki:2016ehc} for a recent review).  
The formulation of the BH and DVCS cross section for various polarizations 
is written in these approaches in terms of finite sums of Fourier harmonics in the azimuthal angle, $\phi$,  ``whose maximal frequencies are defined by the rank of the corresponding leptonic tensor in the incoming lepton momentum" \cite{Belitsky:2001ns,Belitsky:2005qn}. 

It has however, become desirable to formulate the cross section for deeply virtual exclusive processes within a framework that is on one side more transparent and apt for a direct interpretation of the data and that, on the other side, is general and comprehensive of all of the aforementioned processes. 
This goal is even more urgent since a new era of precision data taking is about to start at both Jefferson Lab at 12 GeV, and at the COMPASS experiment at CERN, with the future perspective of an Electron Ion Collider (EIC) just around the corner \cite{Deshpande:2016goi,Ent:2016lod}.
\begin{figure}
\begin{center}
\includegraphics[width=8cm]{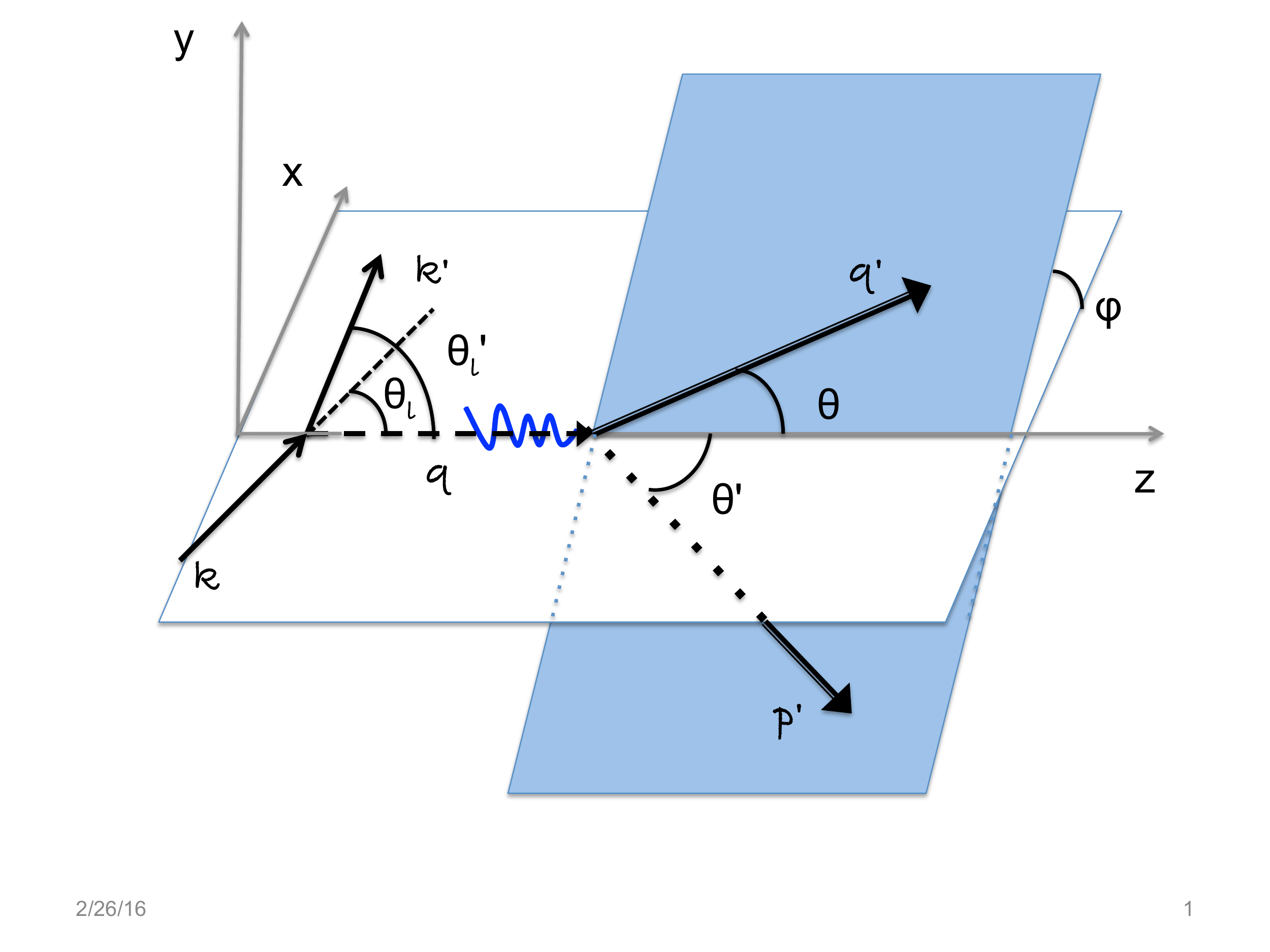}
\caption{Kinematic setting for the exclusive DVCS process.}
\label{fig:kinematics}
\end{center}
\end{figure}

Here we presented some highlights pertaining to the reformulation of the cross section decomposition directly in terms of helicity amplitudes for both BH and DVCS. We discussed the simple case of the unpolarized cross section as an example. The complete treatment of all polarization configurations in DVCS as well as in TCS and DDVCS can be treated along the lines of the given example, and will be discussed in forthcoming papers. 
An important feature of the new formulation is that it both clarifies and it allows one to control the scale and target mass dependence appearing in the BH and DVCS contributions.  

An immediate outcome of our work is that we can single out the twist three GPDs that allow us to directly measure a so far unmeasured contribution in the spin puzzle, namely the Orbital Angular Momentum (OAM) component of the nucleon
\cite{Penttinen:2000dg,Kiptily:2002nx,Rajan:2016tlg,Courtoy:2013oaa,Hatta:2012cs}.  
Looking to the future, the helicity amplitudes formalism is a necessary step for defining processes sensitive to GTMDs, where more than one azimuthal angle is needed.

\item \textbf{Acknowledgments}
 I thank my collaborators on this project, Gary Goldstein, Osvaldo Gonzalez Hernandez and Abha Rajan. I also thank Dustin Keller for guidance and discussions on the MC implementation of our results.
 This work was supported by the U.S.~DOE 
 through the grants 
\#DE-SC0016286 and DOE Office
of Science, Office of Nuclear Physics, through the TMD
Topical Collaboration.

\end{enumerate}


\newpage
\subsection{Time-like Compton Scattering with transversely polarized target}
\addtocontents{toc}{\hspace{2cm}{\sl V.~Tadevosyan}\par}
\setcounter{figure}{0}
\setcounter{table}{0}
\setcounter{equation}{0}
\setcounter{footnote}{0}
\halign{#\hfil&\quad#\hfil\cr
\large{Vardan Tadevosyan\footnote{Email:  tadevosn@jlab.org}}\cr
\textit{ANSL(YerPhysI)}\cr
\textit{Physics Department}\cr
\textit{Armenia}\cr}

\begin{abstract}
A project under development for TCS measurement with transversely polarized target in Hall C at JLab was presented. Physics case and motivation, proposed experimental setup, simulation results and outlooks were outlined.
\end{abstract}

\begin{enumerate}
\item \textbf{Introduction}
TCS is closely related to DVCS, measurement of which is considered the most convenient way of studying GPDs. The two are limiting cases of double DVCS, with hard scale provided by virtuality of incoming (DVCS) or outgoing (TCS) photons. In that sense TCS is inverse of DVCS, with amplitudes complex conjugate at LO of $\alpha_s$ and leading twist, and CFFs same as for DVCS. On the way of obtaining GPDs, combining experimental data from TCS and DVCS measurements may reduce uncertainties on fits to CFFs.

TCS is not attainable alone but in interference with Bethe-Heitler process. While at JLab kinematics the TCS cross section is much smaller than for BH, contribution of TCS in interference with BH is quite significant ($\sim$30\%). While BH alone does not yield target single spin asymmetry because of purely real amplitude, interference with TCS may produce quite measurable asymmetries up to 20\%, as evident from presented model calculations. Particularly, the asymmetry is sensitive to the GPD $E$, which bears information on quark angular momentum. Predicted beam-target double spin asymmetries are also significant, though more complicated for analysis.

This project is focused on the measurement of spin asymmetries with transversely polarized target, which will allow for access to imaginary parts of the $\tilde{H}$  and $E$ GPDs. Double spin asymmetry with transversely polarized target and circularly polarized photon beam (sensitive to real part of the TCS amplitude) will also be measured, provided the polarized photon beam available. The measurement will be complementary to CLAS12 E12-12-001 and SoLID E12-12-006A proposals, which are to measure TCS cross section and beam asymmetry with unpolarized target. The project also envisages measurement of the TCS cross section for cross check with E12-12-001 (sensitive to real parts of CFFs).

The proposed experimental setup consists of the UVA polarized ammonia target, pairs of X and Y trackers, X and Y hodoscopes and electromagnetic calorimeters. To have target polarization perpendicular to the beam direction, the target setup will be rotated by 90$^\circ$ with respect to vertical axis. In this configuration the angular acceptance will be restricted by the target's magnet coils and constructive elements of the target setup to $\pm$ 17$^\circ$ horizontally and $\pm$ 26.5$^\circ$ vertically.

Positioned close to exit windows of the scattering chamber trackers are intended for coordinate measurements and for providing start time for the TOF system. They will be assembled from 1 mm radiation resistant scintillation fibers and will ensure $\sim$0.9 mm space resolution. The multi-anode phototubes for light detection will be moved away from the high magnetic field area, and light to them will be delivered via $\sim$2.5 m long wave-length shifters.

The 1 cm thick scintillator hodoscopes for detection of recoil protons will be positioned at $\sim$1 m away from trackers. Protons of up to $\sim$1 GeV/c momenta will be identified by combining TOF (expected time resolution $\sim$200 ps) and $dE/dX$ signal.

The decay pair of leptons will be detected in a pair of lead tungstate calorimeters positioned just behind the hodoscopes. These calorimeters will be clone of the NPS electromagnetic calorimeter, albeit the transverse sizes different. $\sim$1500 lead tungstate blocks of 2x2 cm$^2$ in cross section will be needed for full coverage of angular acceptance. Expected energy and coordinate resolutions are 2.45\% and 2.7 mm respectively at 1 GeV energy (like in a similar PRIMEX HYCAL calorimeter).

Acceptance studies are done by sampling Bethe-Heitler events at target and tracking the recoil proton and decay leptons through the detector setup. Presented distributions of kinematic quantities show the BH maxima, where TCS is suppressed, are out of acceptance. $\theta_{CM}$ ranges from 30$^\circ$ to 140$\circ$ and peaks at right angle. $\phi_{CM}$is centered around 0$^\circ$ and 180$\circ$. A 3D phase space volume covering $Q^{\prime 2}$ from 4 to 9 GeV$^2$, $\xi$ from 0.1 to 0.3 and $-t$ from 0.1 to 1 GeV$^2$ can be divided in 7 regions for TCS studies.

Currently steps are taken to develop the project to a full proposal. The collaboration is interested in modifications of the UVA target in transverse polarization mode aimed at increase of angular acceptance (B.Wojtsekhowski). The project would greatly benefit from High Energy Photon Source, which allows for significant increase of merits of the projected experiment.


\end{enumerate}

%

\newpage
\subsection{Short Range Correlations and Hard Processes}
\addtocontents{toc}{\hspace{2cm}{\sl M.~Sargsian}\par}
\setcounter{figure}{0}
\setcounter{table}{0}
\setcounter{equation}{0}
\setcounter{footnote}{0}
\halign{#\hfil&\quad#\hfil\cr
\large{Misak Sargsian \footnote{Email:  sargsian@fiu.edu}}\cr
\textit{Florida International University}\cr
\textit{Physics Department}\cr
\textit{Miami, Florida, U.S.A.}\cr}

\begin{abstract}
The potential of the reaction $\gamma + A \rightarrow N^\prime + \pi + (N_r) + X$ for studying outstanding issues of short-range correlations is presented. 
\end{abstract}

\begin{enumerate}
\item \textbf{Introduction}
One of the main conditions for probing short-range nuclear structure is to provide large momentum transfer to the nucleus 
which allows an instantaneous removal of the correlated nucleon from the ground state of the nucleus.
In the case of the photon beam the relevant elementary hard process can be the pion photo-production at large center of mass angles of 
$\gamma + N$ scattering.   
For this we consider the reaction:
\begin{equation}
\gamma + A \rightarrow N^\prime + \pi + (N_r) + X
\label{reaction}
\end{equation}
where $N^\prime$ and $\pi$ are the products of elementary $\gamma + N_{bound} \rightarrow N^{\prime} + \pi$ reaction. Here the additional
nucleon, $N_r$ can be detected in the recoil kinematics corresponding to the production of spectator nucleon from the 2N or 3NN SRCs.
The fixed large center of mass angle provides large sensitivity of the elementary process to the high momentum component of 
nuclear wave function since the cross section scales as  ${d\sigma\over dt} \sim s^{-7}$ and prefers the scattering from the nucleon moving along  
the beam that minimizes the invariant energy $s$.  

\item \textbf{Probing outstanding issues of Short Range Correlations}
The reaction (\ref{reaction}) can be used for systematic studies of the following outstanding issues of SRCs:\\
{\bf - Deuteron Structure in $500-800$~MeV/c region:} There is an important issue about the relative strength of the high momentum 
component of the deuteron. The predictions of current theoretical models based on phenomenological potentials, one-boson exchange models 
and effective theories are widely disagree.  Currently the deuteron  momentum distribution is probed for up to $550$~MeV/c and its extension
is essential for clarifying this issue;\\ 
{\bf - Mapping out the kinetic energy profile of the proton in asymmetric nuclei:} Recent studies of isospin structure of NN SRCs indicate on 
the existence of the strong momentum sharing in asymmetric nuclei, which results in protons having  larger proportion of 
high momentum component.  The intriguing question is whether the average kinetic energy of protons exceeds that of the neutrons for 
neutron rich nuclei.  This question has significant importance  for the equation of state of the dense asymmetric nuclear matter relevant to 
the neutron stars;\\
{\bf - Probing the deuteron in $> 800$~MeV/c region:} The deuteron above the inelastic threshold  of the iso-singlet NN systems represents 
the completely unchartered territory of nuclear forces where one expects the repulsive core to play a dominant role.  Reaction (\ref{reaction}) can 
provide unique access to this domain and allow us to perform an important  measurements complementary to the 
exclusive deuteron electro-disintegration experiments planned at Jefferson Lab for  
the same domain of momenta;\\
{\bf - Observation and systematic studies of 3N SRCs:} Currently the only experiment dedicated to the studies of 3N SRCs is the 
measurement of inclusive $A(e,e^\prime)X$
reactions at $x>2$. For comprehensive investigation one needs also semi-inclusive processes in which the products of 3N SRCs are detected and 
their angular correlations are observed. Such   measurements can be performed using reaction (\ref{reaction}) in which the recoil 
nucleon is detected in the kinematics in which the momentum fraction $\alpha >2$.

\medskip

{\bf Hard Photodisintegration Processes:} Hard photo-disintegtraion processes at large center of mass angles are one of the few processes for 
which there were an observation  of the quark-counting rule relevant to the nuclear targets.  There were several experimental 
studies performed at Jefferson Lab which extended the observation  of the $s^{-11}$ scaling of  the 
deuteron break-up into the $pn$ pair at large center of mass angles  for up to $5$~GeV energies of photon beam.  
There were also two experiments measuring the several polarization observables of the  deuteron break-up reaction, with one 
of the main results being the observation of the large magnitude of transferred longitudinal polarization.

The similar experiments  studying the $^3He$ break into the hard $pp$ and soft $n$ observed tantalizing signatures of the similar scaling with the profile of 
the energy dependence  similar to that of the hard $pp$ scattering. If confirmed, the latter will provide a significant tool for 
studying the QCD structure of NN interaction using photodisintegration processes.

Finally, the recent measurements  of the large center of mass angle  $^3He$ break-up to the $pd$ pair resulted in the surprising 
observation of $s ^{-17}$ scaling in agreement with the quark counting rules.    

All these indicate that hard photodisintegration 
reactions involving light nuclei represent a unique testing grounds for  nuclear QCD processes. In fact the  possibility 
of producing  high intensity photon beams  will make JLab  the only laboratory  were such processes can be studied.


\end{enumerate}

%

\newpage
\subsection{Transparency studies in  large angle exclusive    $\gamma  A\to \mbox{meson + baryon + }A^* $reactions}
\addtocontents{toc}{\hspace{2cm}{\sl M.~Strikman}\par}
\setcounter{figure}{0}
\setcounter{table}{0}
\setcounter{equation}{0}
\setcounter{footnote}{0}
\halign{#\hfil&\quad#\hfil\cr
\large{Mark Strikman \footnote{Email:  mxs43@psu.edu}}\cr
\textit{the Pennsylvania State University}\cr
\textit{Physics Department}\cr
\textit{State College, PA 16802, U.S.A.}\cr\cr
\large{A.B. Larionov }\cr
\textit{Institut f\"ur Theoretische Physik, Universit\"at Giessen, D-35392 Giessen, Germany}\cr
\textit{National Research Centre "Kurchatov Institute", 
             123182 Moscow, Russia}\cr}

\begin{abstract}
We discuss that studies of the semiexclusive large angle photon-nucleus reactions with tagged photon beams of energies 6 $\div$ 10 GeV which can be performed in Hall D at Thomas Jefferson National Acceleration Facility (TJNAF) would allow to probe several aspects of the QCD dynamics: establish the $t$ -range in which transition from soft to hard dynamics occurs, compare the strength of the interaction of various mesons and baryons with nucleons at the energies of few GeV, as well as look for the color transparency effects.
\end{abstract}

\begin{enumerate}
\item \textbf{Introduction}
Large angle high energy exclusive processes provide an effective tool for probing the short-range structure of nuclei \cite{Farrar:1988mf,Piasetzky:2006ai}. So far such studies were performed using proton and electron  beams In  most  studies  a rather limited statistics was accumulated.
Complementary studies can be performed at Jlab using photon beams in reactions 
\begin{equation}
  \gamma + A \to h_1 +h_2 + (A-1)^*~.
\label{aqu}
\end{equation}
Such studies would allow to  check validity of factorization of the cross section the product of  the decay (spectral) function, elementary cross section and the absorption factor. They will also   be sensitive to the EMC like effects due to break down of the many nucleon approximation for the nucleus wave function in the regime where the hard probe - nucleon interaction is dominated by scattering in point-like configurations \cite{Frankfurt:1988nt}.  

Hence understanding of the interaction dynamics is critical for the program of study of the short - range correlations using reaction (1).
It was pointed out in  \cite{Larionov:2016mim} that these reactions allow also  to probe several aspects of the QCD dynamics:
establish the $t$-range in which transition from soft to hard dynamics occurs, compare the strength of the interaction of various mesons and baryons with nucleons
at the energies of few GeV, as well as look for the color transparency effects. 

Experimentally the reactions $\gamma N \to "meson" N$ follow the expectations of the quark counting rules at $\theta_{cm}=90^o$ \cite{Brodsky:1973kr}  which predict that the $s$ dependence of this process ($s^{-7}$)  is slower than  for the case of pion scattering. It is worth noting here that this statement is more solid than determination of the power itself as one can use more complicated parametrizations of the cross section in the limited $s$ range to fit the $\gamma N$ data to a different power.

It would be desirable to study the reaction (1) in three kinematic regions where different physics dominates.

\begin{itemize}
\item{ In the  t-range  $2 >  -t  >  1 \mbox{GeV}^2$   geometric, Glauber - like dynamics is expected to dominate. It would be possible to address the question of the validity of the vector dominance model (VDM) approximation, compare strange and non strange channels. One would need to select sufficiently large $s$ to avoid dominance of the processes of  exclusive production of baryon resonances.   }
\item {For  $-t  > 3 \div 2 \mbox{GeV}^2$ one is expected to reach the photon transparency
- transition from  regime of VDM hadron-like   unresolved photon to the regime where photon is  acting as an elementary  (point-like) particle.
For $\theta_{cm}=90^o$ this regime may start right above the resonance region leaving little $s$ range for the Glauber like regime. Hence covering a wide range of the c.m. angles in the planned studies is very important.}

\item {The   regime of photon  transparency  can be used also  for comparing strength of interactions of mesons ($\pi, \rho, \eta, \eta', K^*$), and  baryons $(N,\Delta)$ using processes (1).   The optimal quantity to study would be the double ratio 
\begin{equation} 
R(A) = {\sigma(\gamma +A\to h_1+N + (A-1)^*)\over \sigma(\gamma +A\to h_2+N + (A-1)^*)}
/ {\sigma(\gamma +N\to h_1+N)\over \sigma(\gamma +N\to h_2+N)},
\end{equation}
which is very sensitive to $ \sigma_{in}(h_1 N)/\sigma_{in}(h_2 N)$,  see Fig.~3 in \cite{Larionov:2016mim}. }

\item {At large $ -t  \ge  4 \div 3 \mbox{GeV}^2$
it would be possible to explore the  onset of the color transparency regime. Such a study would help to resolve old puzzles of the transparency measurements with proton beam, and observe  point-like configurations in hadrons. It would link in a natural way to the previous and future studies of color transparency at Jlab in the 
 $(e,ep), (e,e\pi)$ reactions, see review in 
\cite{Dutta:2012ii}.}
\end{itemize}

Numerical studies performed in  \cite{Larionov:2016mim} have found that the nuclear  transparency in  the discussed three regimes is significantly different already for light nuclei and has a very different A-dependence, see Fig. 1. In particular, photon transparency leads to a very large increase of transparency for heavy nuclei.

In conclusion, the discussed reactions have a strong potential for  discovering  new features of the QCD dynamics at intermediate energies. Further topics to be studied include spin effects in the initial state (photo polarization) and in the final state ($\rho, \Delta$).  Studying short-range correlations in the different transparency regimes would give an important additional information about short-range nuclear dynamics.

\begin{figure}
\begin{center}
   \includegraphics[scale = 0.7]{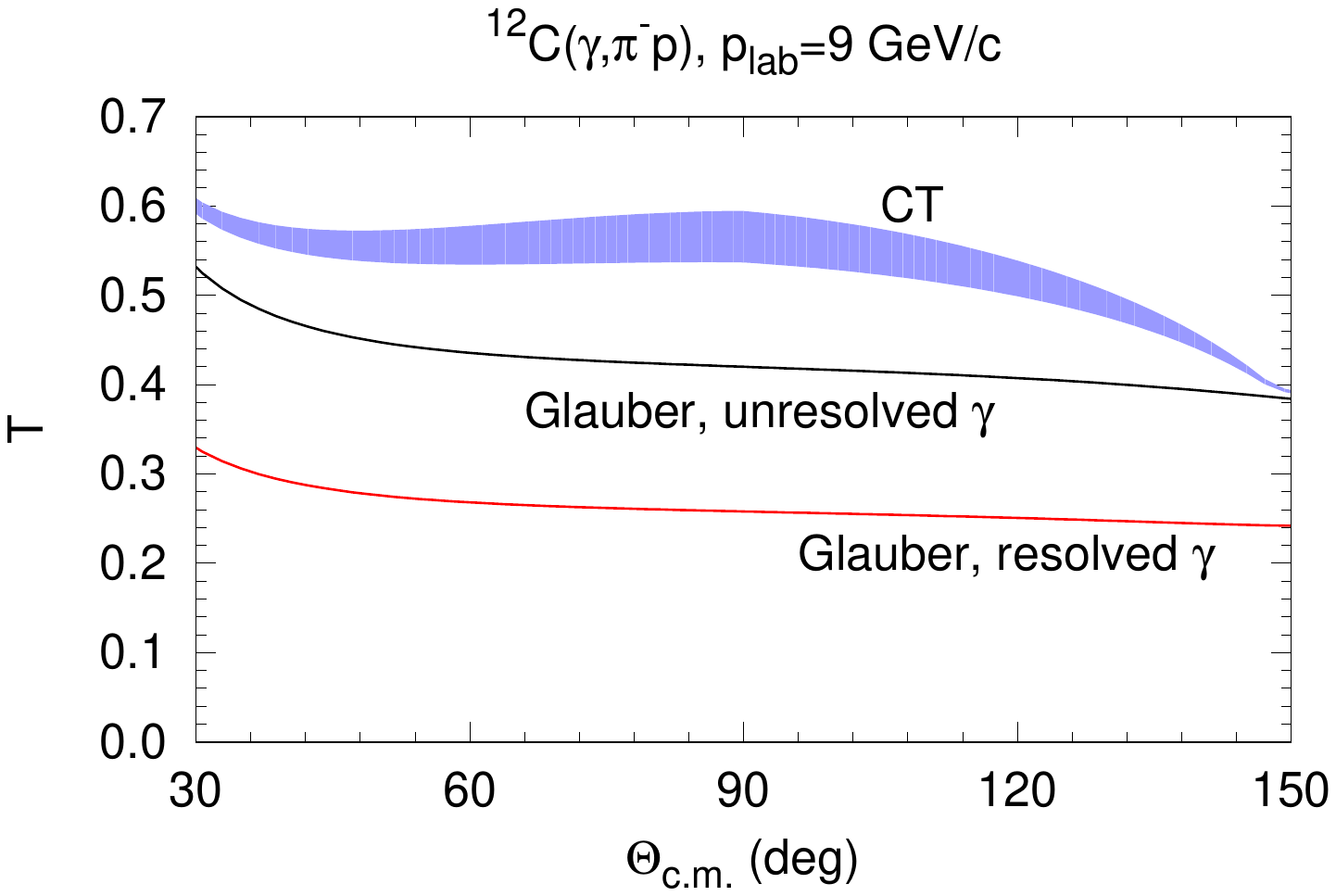}
   \includegraphics[scale = 0.7]{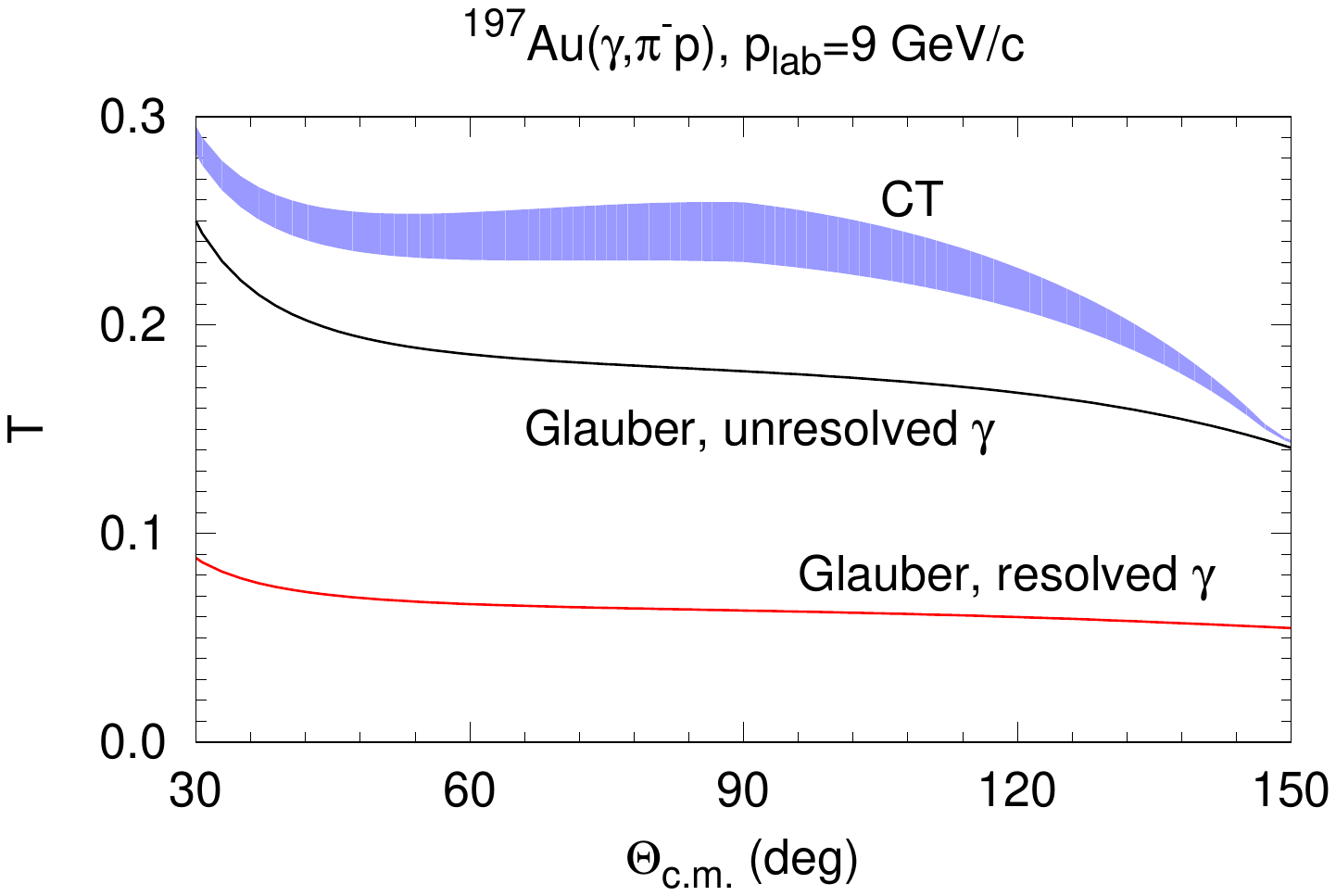}
\end{center}
\caption{\label{fig:photran_t-2gev2} Transparency for the $A(\gamma,\pi^-p)$ semiexclusive process for $^{12}$C and $^{197}$Au target nuclei   
under assumption  of the dominance of the unresolved photon ($\sigma^{\text{eff}}_{\gamma\text{N}}=0$, solid curve) and resolved photon interacting in configuration with 
$\sigma^{\text{eff}}_{\gamma\text{N}}=\sigma_{\rm in}(\pi N)$ (dashed curve). Calculations are performed at $t=-2$ GeV$^2$.}
\end{figure}

\item \textbf{Acknowledgments}
A.L. acknowledges financial support by the Deutsche Forschungsgemeinschaft
(DFG) under Grant No. Le439/9 and the Helmholtz International Center (HIC)
for FAIR.
M.S.'s research was supported by the US Department of Energy Office of Science, 
Office of Nuclear Physics under Award No.  DE-FG02-93ER40771. 

\end{enumerate}


\newpage
\subsection{Compact Photon Source Conceptual Design}
\addtocontents{toc}{\hspace{2cm}{\sl P.~Degtyarenko}\par}
\setcounter{figure}{0}
\setcounter{table}{0}
\setcounter{equation}{0}
\setcounter{footnote}{0}
\halign{#\hfil&\quad#\hfil\cr
\large{Pavel Degtyarenko}\cr
\textit{Thomas Jefferson National Accelerator Facility}\cr
\textit{Newport News, VA 23606, U.S.A.}\cr}

\begin{abstract}
We describe options for the production of an intense photon
beam at the CEBAF Hall~D Tagger facility, needed for creating a 
high-quality secondary $K^0_L$ beam delivered to the Hall D detector. 
The conceptual design for the Compact Photon Source (CPS) apparatus
is presented.
\end{abstract}

\begin{enumerate}
\item \textbf{Introduction}

An intense high energy gamma source is a pre-requisite for the
production of the $K_L^0$ beams needed for the new proposed
experiments at Hall~D~\cite{PDBW}. Here we describe a new approach 
to designing such photon sources. Possible practical implementation, 
adjusted to the parameters and limitations of the available 
infrastructure is discussed. The vertical cut of the Compact Photon 
Source (CPS) model design, and the plan view of the present Tagger 
vault area with CPS installed are shown in Fig.~\ref{fig:CPS}.
%
%
\begin{figure}[!ht] 
\centering 
  \subfloat[Vertical cut plane of the GEANT3 model of the CPS]{%
    \includegraphics[width=0.5\textwidth]{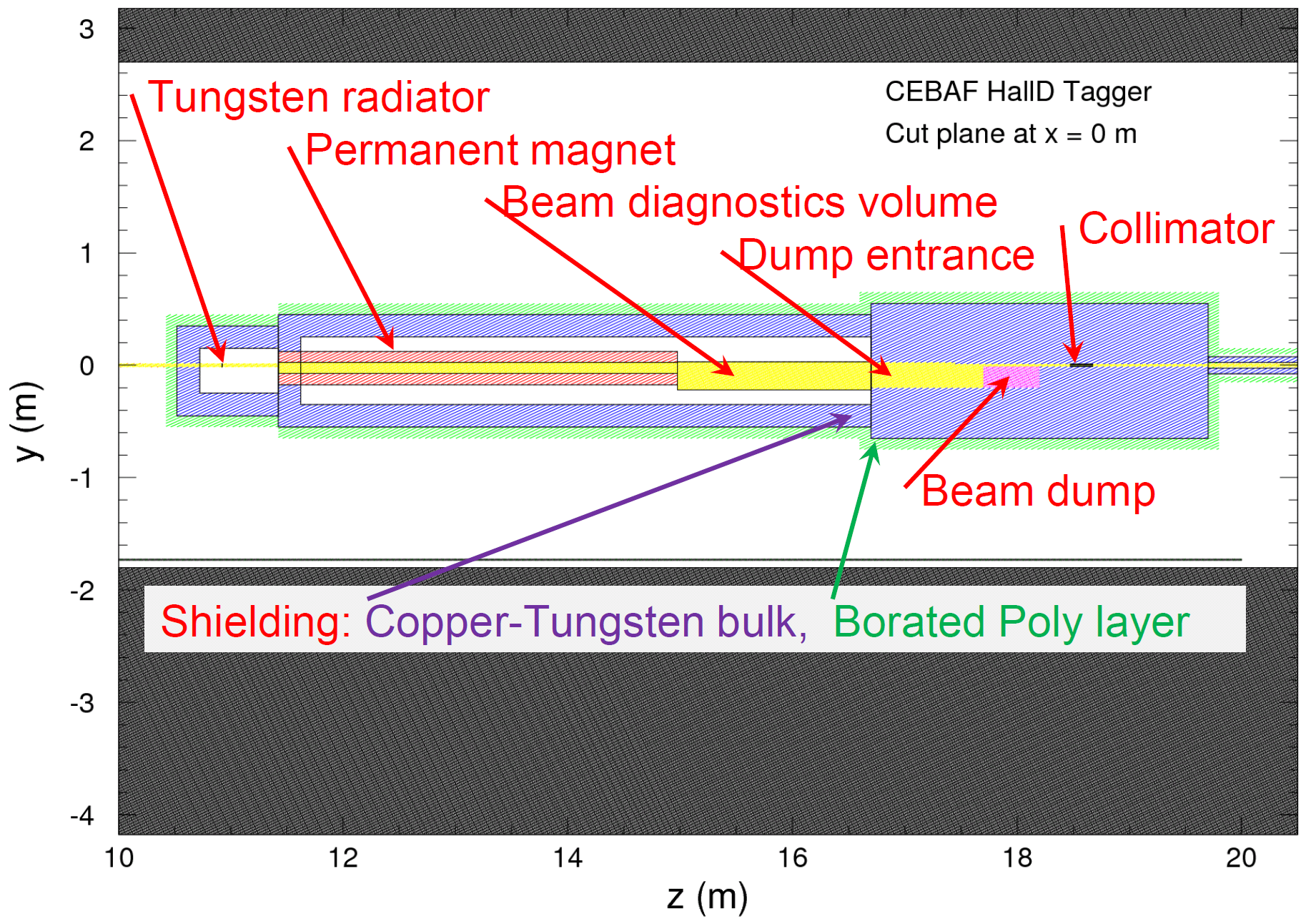} }
  \subfloat[A plane cut of the Tagger vault model]{%
    \includegraphics[width=0.5\textwidth]{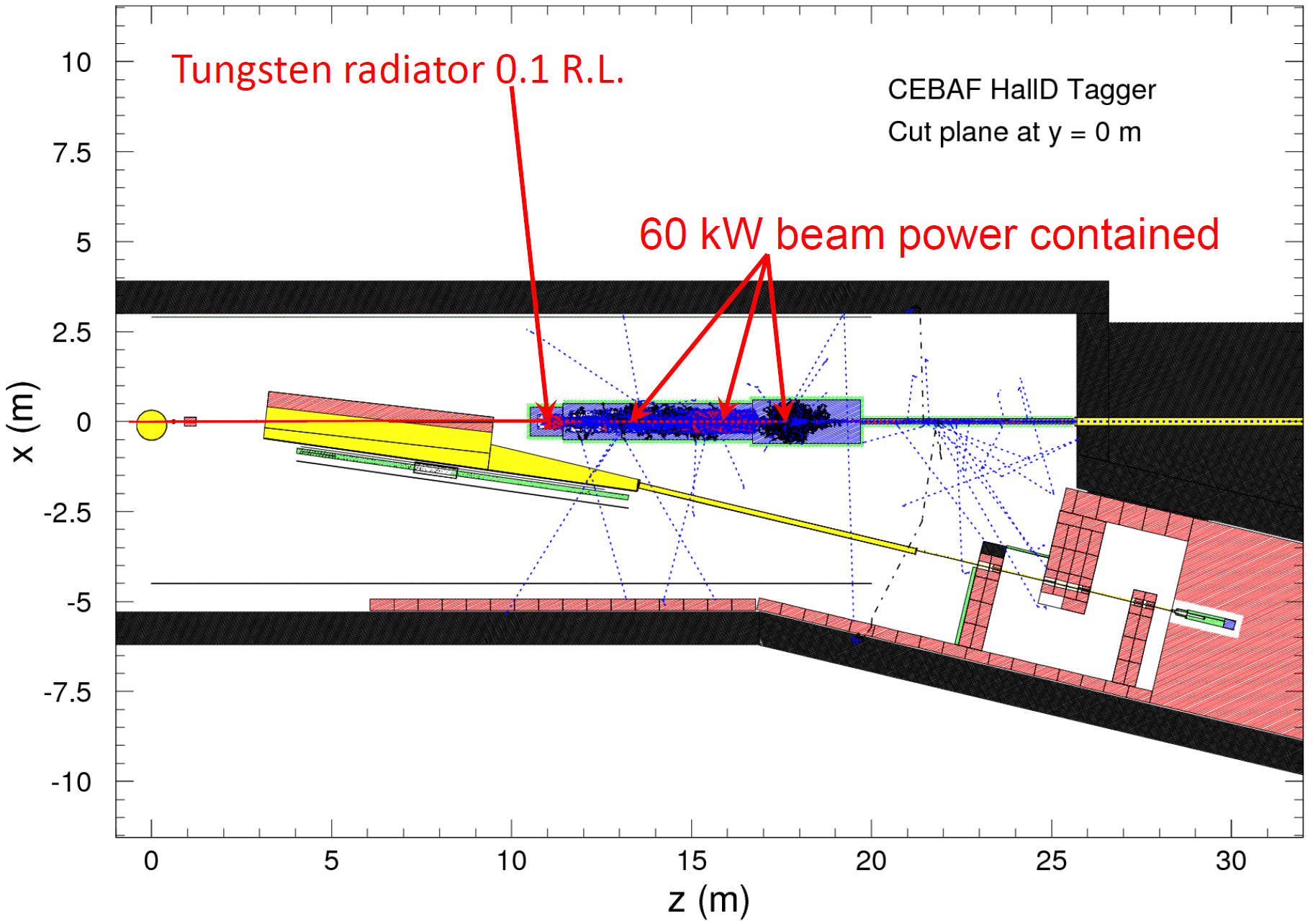} }
  \caption{ Elements of the design are indicated in the plot in panel
    (a). Panel (b) shows the CPS assembly in the Tagger vault and the
    simulation of 2000 beam electrons at 12~GeV. }
\label{fig:CPS}
\end{figure}
%
%

The new design combines in a single properly shielded assembly all
elements necessary for the production of the intense photon beam, such
that the overall dimensions of the setup are limited and the
operational radiation dose rates around it are acceptable  Compared
to the alternative, the proposed CPS solution presents several
advantages, including much lower radiation levels, both prompt and
post-operational due to the beam line elements' radioactivation at the
vault; much less disturbance of the available infrastructure at the
Tagger Area and better flexibility in achieving high-intensity photon
beam delivery to Hall~D. The new CPS solution will satisfy proposed
$K_L^0$ beam production parameters; we do not envision big technical
or organizational difficulties in the implementation of the conceptual
design.

Conceptually there is no problem with the "skyshine" radiation from the
Tagger vault in Hall~D with the new proposed setup. The difference
between the Halls~A \& C and the Tagger vault is in the thickness of the
concrete and berm covering the enclosures. As Halls~A \& C are huge, it's
impossible to make the roofs thick, so they are "as thick as reasonably
possible". They perform as shielding only to a limit. Tagger vault is
smaller and narrower, so the roof there is much thicker. One cannot
remember exactly, and do not have a quick access to the drawings right
now, but one can think it's of the order of 5~m, about twice that of
Halls~A \& C, and it's a lot of shielding. That roof has been designed
to hold the 60~kW local beam dump in it, assuming the proper local
shielding is set around the dump. The dump is there behind the labyrinth
walls, surrounded by about 2~m of iron shielding, using the standard
iron shielding blocks that were available at that time. About the same
thickness of the roof is above it. That's an approved and built solution.

The new CPS dump setup for Hall~D is essentially the same: 60~kW new dump
with optimized high density shielding around it, under the same roof.
Using high density shielding material and making it compact actually saves
the total weight of the shielding, compared to iron. Qualitatively, if you
need a sphere of iron (8~g/cm$^3$) of 2~m radius for the shielding, it
may be roughly replaced by a sphere of 1~m radius made of tungsten-copper
(16~g/cm$^3$), with its weight actually four times smaller. Of course
it's still a lot of weight and it will require engineering analysis for
the building. Our point is that the comparable conditions are already
there, and thus we do not see conceptual problems with the new setup.
Optimization and implementation of the new CPS for Hall~D will of course
require more detailed answers to all these valid questions by everyone
involved.


\end{enumerate}



%
%





\newpage
\subsection{Summary of Optimized Photon Source and Science Opportunities}
\addtocontents{toc}{\hspace{2cm}{\sl T.~Horn, C.~Keppel, C.~Munoz-Camacho, I.~Strakovsky}\par}
\setcounter{figure}{0}
\setcounter{table}{0}
\setcounter{equation}{0}
\setcounter{footnote}{0}
\halign{#\hfil&\quad#\hfil\cr
\large{Tanja Horn \footnote{Email:  hornt@cua.edu}}\cr
\textit{Catholic University of America}\cr
\textit{Washington DC, U.S.A.}\cr
\textit{and Thomas Jefferson National Accelerator Facility}\cr
\textit{Newport News, VA, U.S.A.}\cr\cr
\large{Cynthia Keppel \footnote{Email:  keppel@jlab.org}}\cr
\textit{Thomas Jefferson National Accelerator Facility}\cr
\textit{Newport News, VA 23606, U.S.A.}\cr\cr
\large{Carlos Munoz-Camacho \footnote{Email:  munoz@jlab.org}}\cr
\textit{IPN-Orsay, CNRS/IN2P3}\cr
\textit{France}\cr\cr
\large{Igor Strakovsky \footnote{Email:  igor@gwu.edu}}\cr
\textit{George Washington University}\cr
\textit{Washington DC, U.S.A.}\cr}

\vspace{1cm}
After the scientific presentations were complete, a summary talk was given by C.~Keppel and a disussion of the path forward was held. The two main topics of the discussion were:
\begin{itemize}
\item{New Opportunities with High-Intensity Photon Sources}
\item{Summary of Optimized Photon Source and Science Opportunities}
\end{itemize}

A possible photon source would give a gain in figure-or-merit of a factor of 30 for some experiments. For processes such as wide-angle compton scattering and timelike compton scattering the high intensity photon source could be coupled with high-precision calorimetry. 

\begin{enumerate}
\item \textbf{New Opportunities with High-Intensity Photon Sources}

A high-intensity photon source provides a wide array of science opportunities. These include Wide-Angle Compton Scattering (WACS), the search for missing hyperons, which can be connected to QCD thermodynamics of the early universe, as well as WACS exclusive photoproduction, Wide-Angle meson production, Timelike Compton Scattering, Short Range Correlations, and photoproduction of few body systems.

The process of WACS is one of the least well understood processes in hadronic physics and is very compelling. Discussions identified the need for a cohesive approach in both scientific focus and experiment design. A clear priority for WACS are measurements at high values of $s$ and $t$ and $u$. Scans in $s$ for center of mass angles of 90 and $\sim$ 120 degrees are of interest. Discussion also focused on combining efforts and collaboration to determine the best photon source design, to evaluate the importance of a large acceptance, and a strategy to determine the Hall preference. 

The intense photon source is one component of the $K_{L}$ beam. The experimental method can be summarized as follows: electrons hit a tungsten radiator, the resulting photons hit a Be target, a beam to $K_L$ is produced. The search for missing hyperons is a strong motivation for this setup. The large photon intensity and low neutron/$K_L$ flux ratio compared to proton beams are solid arguments. The discussions identified three suggestions to refine the science case: 1) improve the resolution by using 50 ps timing rather than 250 ps and to go to $W<$3 GeV. 2) Consider unpolarized deuterium running with a split of $H/D$ to be determined, 3) include the expected data in Partial Wave Analysis to show the potential impact of new measurements including $p$ and $n$ mix.

\item \textbf{Optimized Photon Source and Science Opportunities}

The requirements of a new optimized photon source include: pure photons, highest energy, high intensity to provide an increase of FOM by factor of 30 in combination with polarized target with reduced heat load, higher polarization and less depolarization. THe main approaches are driven by the Hall A/C WACS efforts. Both design concepts were discussed at the workshop. The Hermetic Compact Photon Source uses a bend magnet that also functions as a beam dump. The small bore allows for a high field. The compact design allows it to be placed close to the target. The Separated Function Dipole and Dump design features a beam dump separated and far away from the target area. It requires more magnet strength to bend the beam. Its design principle reduces activation and radiation to facilitate work on the pivot. In the summary a comparison was made between the two designs based on physics (pure photons, high intensity, high energy), practical (magnetic field, hall integration, cooling needs, inter-hall compatibility, cost), and radiation (minimization of radiation at pivot, minimize dose at de-installation source magnet and dump, minimize dose to hall equipment) considerations. 

A design path forward for the intense photon source emerged based on discussions. It will include simulation benchmarking to establish a common setup for simulations. Simulations should estimate radiological conditions for beam on and off (radiation due to activation) and both neutron and photons. It will also include common numbers/locations for design goals such as $<$2 mrem/hr required at the pivot right after beam off, dose rates at dump and along the beam line, and dose and activation values at specific locations of hall equipment, e.g., the SHMS magnets. Furthermore, the path forward will consider experiments beyond WACS and even beyond Halls A/C. The implementation of goals and timeline will be determined through a series of meetings with the representatives of the photon designs and WACS, the $K_L$ effort, the NPS collaboration and Jefferson Lab.

\end{enumerate}

\newpage
\section{List of Participants of HIPS2017 Workshop}

\begin{itemize}	
\item Salina Ali, CUA       				  <95ali@cua.edu>
\item Lee Allison, ODU       		 		 <salli008@odu.edu>
\item Moskov Amaryan, ODU       			  <mamaryan@odu.edu>
\item Rakitha Beminiwattha, Louisiana Tech University	<rakithab@latech.edu>
\item Alexandre Camsonne, Jefferson Lab       		<camsonne@jlab.org>
\item Marco Carmignotto, CUA       			<marcoapc@jlab.org>
\item Donal Day, University of Virginia   		<dbd@virginia.edu>
\item Pavel Degtiarenko, Jefferson Lab       		<pavel@jlab.org>
\item Dipangkar Dutta, Mississippi State University   	<d.dutta@msstate.edu>
\item Rolf Ent, Jefferson Lab       		<ent@jlab.org>
\item Jos\'e~L.~Goity, Hampton U./JLab 		  <goity@jlab.org>
\item David Hamilton, University of Glasgow (UK)	 <david.j.hamilton@glasgow.ac.uk>
\item Or Hen, MIT				  <hen@mit.edu>
\item Tanja Horn, CUA/JLab				  <hornt@cua.edu>
\item Charles Hyde, ODU			  	  <chyde@odu.edu>
\item Greg Kalicy, CUA				  <kalicy@cua.edu>
\item Dustin Keller, University of Virginia   		<dustin@jlab.org>
\item Cynthia Keppel, Jefferson Lab       		<keppel@jlab.org>
\item Chan Kim, GWU				  <kimchanwook@gwu.edu>
\item Ed Kinney, University of Colorado		<Edward.Kinney@colorado.edu>
\item Peter Kroll, University of Wuppertal (Germany)	<pkroll@uni-wuppertal.de>
\item Aleixei B. Larionov, University of Giessen (Germany) 
\item Simonetta Liuti, University of Virginia   <sl4y@virginia.edu>
\item Maxim Mai, GWU 			  	  <maximmai@gwu.edu>
\item Arthur Mkrtchyan, CUA				  <mkrtchya@jlab.org>
\item Hamlet Mkrtchyan, ANSL (YerPHI)			<hamlet@jlab.org>
\item Carlos Munoz-Camacho, IPN-Orsay, CNRS/IN2P3 (France)	<munoz@jlab.org>
\item James Napolitano, Temple U.		  <tuf43817@temple.edu>
\item Gabriel Niculescu, James Madison Univ.		  <gabriel@jlab.org>
\item Maria Patsyuk, MIT				  <mpatsyuk@mit.edu>
\item Gonaduwage Perera, University of Virginia   	<darshana@virginia.edu>
\item Hashir Rashad, ODU       		 		 <hashir@odu.edu>
\item Julie Roche, Ohio U.       		 	<rochej@ohio.edu>
\item Misak Sargsian, Florida International U.       	<sargsian@fiu.edu>
\item Simon Sirca, Univ. of Ljubljana (Slovenia)      <simon.sirca@fmf.uni0lj.si>
\item Igor Strakovsky, GWU                        <igor@gwu.edu>
\item Mark Strikman, Penn State                        <mxs43@psu.edu>
\item Vardan Tadevosyan, ANSL (YerPHI)			<tadevosn@jlab.org>
\item Richard Trotta, CUA				  <trotta@cua.edu>
\item Rishabh Uniyal, CUA				  <uniyal@cua.edu>
\item Andres Vargas, CUA				  <vargasa@cua.edu>
\item Bogdan Wojtsekhowski, Jefferson Lab		  <bogdanw@jlab.org>
\item Jixie Zhang, University of Virginia      		<jixie@jlab.org>
\item Aaron Dominguez, CUA				  <domingueza@cua.edu>
\end{itemize}
\end{document}